\documentclass[a4paper]{article} % KOMA-Script article scrartcl
\usepackage{lipsum}
\usepackage{url}
\usepackage{mathpazo}
\usepackage[T1]{fontenc}
\usepackage[utf8]{inputenc}
\usepackage[english]{babel}
\usepackage[left=3cm,top=3cm,right=3cm,bottom=3cm]{geometry}

\usepackage{natbib}
\usepackage{amsmath}
\usepackage{amssymb}
\usepackage{amsthm}
\usepackage{hyperref}
\usepackage{graphicx}
\usepackage{array}% http://ctan.org/pkg/array
\usepackage{longtable}
\usepackage{verbatimbox}
\usepackage{textcomp}
\usepackage{gensymb}
\usepackage{verbatimbox}

\usepackage[title]{appendix}
\graphicspath{{fig/}}
\usepackage{bm}
\usepackage{upgreek}
\usepackage{mathrsfs}
\usepackage{xcolor}
\usepackage{placeins}
\graphicspath{{fig/}}

\usepackage{subcaption}
\usepackage[small, sc]{titlesec}

\usepackage{xcolor}
\hypersetup{
colorlinks,
linkcolor={red!50!black},
citecolor={blue!50!black},
urlcolor={blue!80!black}
}
\theoremstyle{plain} 
\DeclareMathOperator{\diver}{div}
\DeclareMathOperator{\grad}{grad}
\DeclareMathOperator{\Diver}{Div}
\DeclareMathOperator{\Grad}{Grad}

\DeclareMathOperator{\tr}{tr}

\usepackage{mathtools}
\usepackage{pgfplots}
\pgfplotsset{/pgf/number format/use comma,compat=newest}

\renewcommand\epsilon{\varepsilon}
\newcommand{\R}{\mathbb{R}}
\newcommand{\N}{\mathbb{N}}

\newcommand{\vect}[1]{\boldsymbol{#1}}
\newcommand{\tens}[1]{\mathsf{#1}}

\begin{document}
\title{\textsc{Morphomechanical model of the torsional c-looping in the embryonic heart}}

\author{\textsc{G. Bevilacqua}$^1$\thanks{\href{mailto:giulia.bevilacqua@polimi.it}{\texttt{giulia.bevilacqua@polimi.it}}} $\,\,-$
\textsc{P. Ciarletta}$^{1}$\thanks{\href{mailto:pasquale.ciarletta@polimi.it}{\texttt{pasquale.ciarletta@polimi.it}}} $\,\,-$
\textsc{A. Quarteroni}$^{1,2}$\thanks{\href{mailto:alfio.quarteroni@polimi.it}{\texttt{alfio.quarteroni@polimi.it}}}\bigskip\\
\normalsize$^1$ MOX -- Politecnico di Milano, Piazza Leonardo da Vinci 32, Milano, Italy.\\
\normalsize$^2$ Mathematics Institute, \'Ecole Polytechnique F\'ed\'erale de Lausanne,\\ \normalsize Av. Piccard, CH-1015 Lausanne, Switzerland ({\em Professor Emeritus}).}
\date{}

\maketitle

\begin{abstract}
Before septation processes shape its four chambers,  the embryonic heart is  a  straight tube that spontaneously  bends and twists breaking the left-right symmetry. In particular, the heart tube is subjected to a cell remodelling inducing  ventral bending and dextral torsion  during the c-looping phase.  In this work we propose a morphomechanical  model for the torsion of the heart tube, that behaves as a nonlinear elastic body. We hypothesize that this spontaneous looping can be modeled as a mechanical instability due to accumulation of residual stresses induced by the geometrical frustration of tissue remodelling, which mimics the cellular rearrangement within the heart tube. Thus, we perform a linear stability analysis of the resulting nonlinear elastic boundary value problem  to determine the onset of c-looping as a function of the aspect ratios of the tube and of the internal remodelling rate.  We perform numerical simulations to study the fully nonlinear morphological transition, showing that the soft tube develops a realistic  self-contacting looped shape in the physiological range of geometrical parameters.\\

{\em Keywords}: embryogenesis, c-looping, heart tube, elastic stability, remodelling.

\end{abstract}

\section{Introduction}
\label{sec:intro_clooping}

In human embryos, the heart is the first functioning organ. Cardiac 
contractions begin about $17$ days post-conception, when the heart is essentially a single, relatively straight, muscle-wrapped tube \cite{taber1995mechanics}. In the next stage, the heart tube (HT) bends and twists developing a curved shape towards the right side of the embryo. 
Cardiac looping represents the first visible sign of left-right asymmetry in vertebrate embryos. Its inception received much attention in clinics, since spontaneous abortions during the first trimester may occur for cardiac malformations caused by serious structural defects and abnormalities induced by minor looping perturbations \cite{srivastava1997knowing}.
After that looping is complete, the heart reaches the  required configuration for further development into a four-chambered pump.

\begin{figure}[h!]
	\centering
	\includegraphics[width=0.7\textwidth]{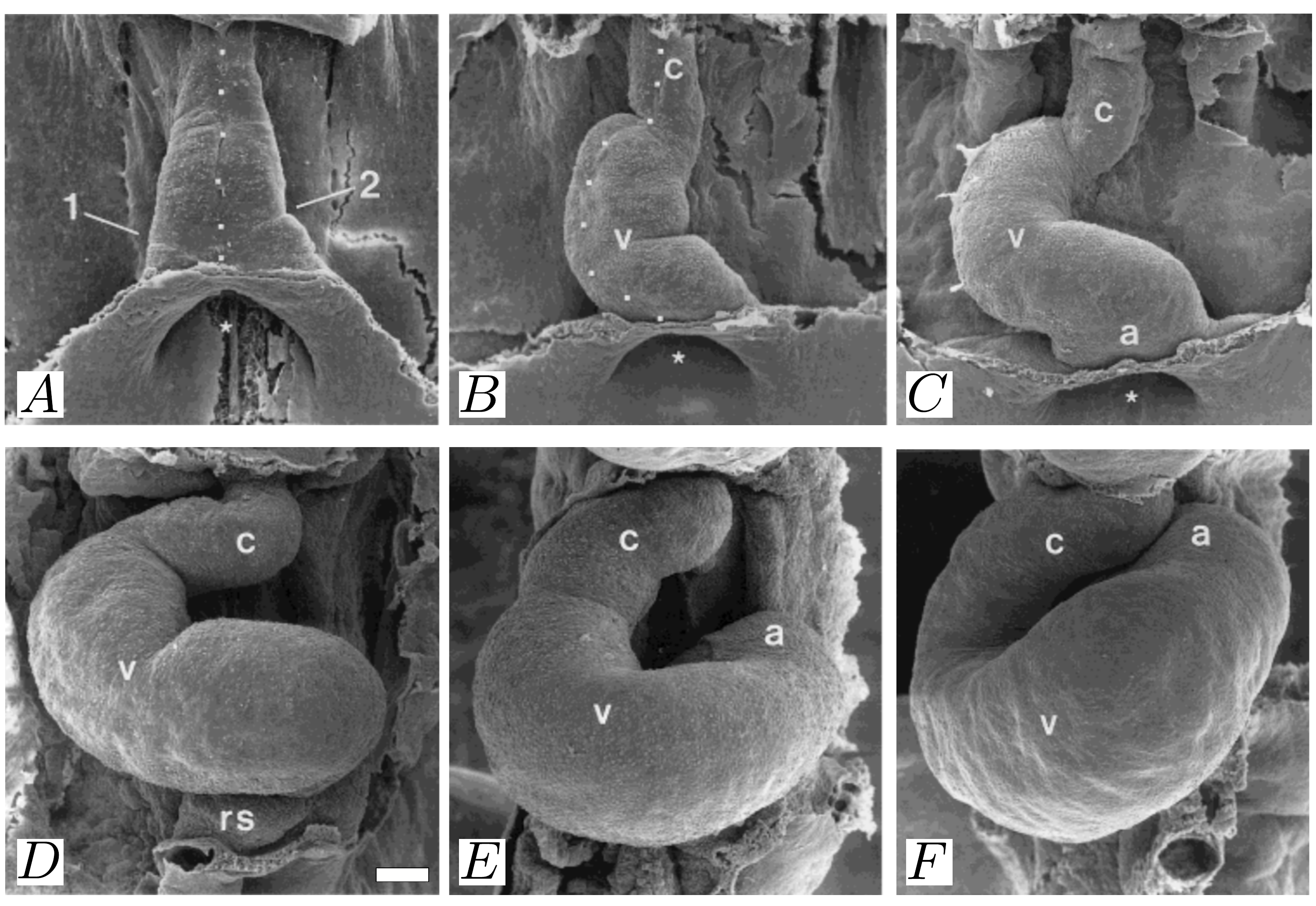}
	\caption{Ventral views showing the positional and morphological changes of the embryonic heart tube between its first morphological appearance at Hamburger Hamilton (HH)-stage 10 \cite{hamburger1951series} (A), the beginning of the c-looping, HH-stage 12/13 \cite{hamburger1951series} (C), the end of the phase of dextral-looping at HH-stage 13 \cite{hamburger1951series} and the transformation from the c-shaped loop into the s-shaped one (D)-(F). [c], primitive conus; [v], primitive ventricular bend; [a], primitive atria.  Scale bar 100${\rm \mu m}$.  Figures adapted from  \cite{manner2000cardiac}.}
	\label{fig:exp_result}
\end{figure}

Collecting well-defined images of the human embryo in its first days of formation is a hard procedure due to the necessity to avoid invasive procedures on the mother's body and to its very small size \cite{voronov2002cardiac}. In order to circumvent these difficulties, researchers used chick embryos for studying cardiac morphogenesis. In fact, the development of the chick heart has the same characteristic duration as the human one: the process takes $21$ days and it can be divided into $46$ stages, characterizing the morphological changes \cite{hamburger1951series}. Moreover, the chick embryo can be cultured both {\em in-vivo} and {\em in-vitro} to better understand the underlying biological and physical mechanisms \cite{dehaan1967development, voronov2002cardiac, taber2010role}. 
Looping begins at stage $10$ and consists of two main phases: \textit{c-looping} and \textit{s-looping} \cite{patten2008early, manner2000cardiac}. During normal c-looping (stages $9-12$), the heart tube transforms from a straight tube into a c-shaped one,  see (C) of Fig. \ref{fig:exp_result}, via two main deformations: a
ventral bending and a dextral (rightward) torsion \cite{taber1995mechanics,manner2000cardiac, shi2014bending}.
%\footnote{In the biology literature, \emph{torsion} of the heart tube usually is referred to as \emph{rotation}. The former term is more appropriate from an engineering perspective, 
%however, as “bending” and “torsion” describe deformation, whereas “rotation” describes a rigid-body motion. Because of the constraints at the ends of the heart tube, any rotation of the tube must be accompanied by torsion (twist).} 
Hence, the looped HT looks like a helix \cite{manner2014cardiac} since the original ventral surface of the straight heart tube becomes the outer curvature (convex surface) of the looped heart, while the original dorsal side becomes the inner curvature (concave surface), see (A), (B) and (C) of Fig. \ref{fig:exp_result}.  
During s-looping (stages $12-16$), the primitive ventricle from its post c--loop cranial position moves to its definitive caudal position and induces a shortening of the distance between the conotruncus (outflow tract) and the atrium \cite{ramasubramanian2013role,ramasubramanian2019biomechanics}, see (D), (E) and (F) of Fig. \ref{fig:exp_result}. 
%the atrium moves superior to the ventricle, creating the basic final form of the heart 
During the later stages $18-36$, septation processes divide the tube into four chambers.

In the following, we primarily focus on the c-looping of the HT. Since previous works mostly focused on a biological and genetical aspects \cite{manasek1976heart,itasaki1991actin,voronov2002cardiac,voronov2004role}, the looping remain poorly understood from the mechanical point of view. Indeed, it has been highlighted that internal and external physical forces direct bending and torsion in the embryo \cite{taber2003biophysical,shi2014bending_torsion}. Concerning the HT bending, the main idea is that it is driven by differential growth: cells on the dorsal and ventral surfaces growth primarily by 
hypertrophy (bigger cells) than by hyperplasia (more cells). Hence, on the ventral surface the 
size of the cells is smaller than the ones on the dorsal one \cite{shi2014bending}.
The physical origin of dextral  torsion is openly debated. The HT is covered by a membrane,  called splanchnopleure (SPL), which exerts a pressure on the heart. At the end of the tube there are two veins called
omphalomesenteric vein (OVs), one on the right side and the other on the left.   Before the onset of c-looping, the stress in the OVs is different in its caudal and cranial part, due to the different geometry of the veins and to the migration of precardiac cells from the OVs to the HT \cite{voronov2004role}. At stage 9, when c-looping process starts, the one on the right has a larger diameter than the other one.
This phenomenon represents a break of symmetry in the structure which coupled with the action of the SPL pressure might cause the torsion and the position of the heart in the left part of the body \cite{voronov2002cardiac,voronov2004role,taber2010role,shi2014bending_torsion}. However, there are several candidate mechanisms for symmetry break in such a complex process. For instance, the SPL membrane has been removed in recent experiments \cite{voronov2004role}. The resulting HT rotation is inhibited but not eliminated, proving that looping involves a combination of different morphogenetic mechanisms, some of which can be redundant \cite{taber2003biophysical,taber2010role}.

Motivated by these experimental results, in this work, for the first time, we propose an original contribution to the mathematical modeling of a fundamental process about cardiac shaping: we introduce a morphomechanical  model for the torsion of the HT. We hypothesize that this spontaneous morphological transition can be modeled as a mechanical instability due to accumulation of residual stresses due to the geometrical frustration imposed by tissue remodelling. Precisely, we assume that the HT is a nonlinear elastic, homogeneous, incompressible body that is  subjected to a torsional remodelling, which mimics the cells flux in the HT \cite{shi2014bending}.  We assume that the tube undergoes a finite torsion by accumulating such a geometrical frustration, and we perform a linear stability analysis of the nonlinear elastic boundary value problem  to determine the onset of c-looping as a function of the geometrical parameters of the HT. Then, we perform numerical simulations to study the post-buckling behaviour in the fully nonlinear morphological transition.

%is one of the few universal solutions of nonlinear isotropic incompressible elastic materials \cite{horgan1999simple}, the interest moves in studying torsional instabilities: \cite{green1958stability} proposes an analytical solution for a fixed class of instability modes; \cite{gent2004torsional} studies the evolution of a twisted cylinder into a kink or a knot; \cite{Ciarletta_2014} investigates the onset of wrinkling instabilities in isotropic, incompressible, hyperelastic full cylinders, while \cite{balbi2015helical} studies the behaviour of the hollow ones. Moreover, as regarding the finite elements implementation, there are few results and most of them are not a fully $3$D model \cite{Bigoni_2018,wang2019elastic,du2020tension}.
%In this work we aim to propose a very simple model for the torsion of the heart tube describing it as an hyperelastic, incompressible hollow cylinder.

The paper is organised as follow: in Section \ref{sec:elastic_model_clooping}, we develop the nonlinear elastic model of the HT. In Section \ref{sec:lin_stab_clooping} we perform a linear stability analysis of the radially symmetric solution and in Section \ref{sec:numerical_simulation_clooping} we implement a finite element code to study the post-buckling behaviour. Finally, in Section \ref{sec:conclusion_clooping} we discuss the outcomes of our model together with some concluding remarks.
 
\section{Mathematical model}
\label{sec:elastic_model_clooping}

In this Section, we define the morphomechanical model of the HT, described as a hyperelastic body subjected to torsional remodelling.

\subsection{Kinematics}

Let
\[
\resizebox{0.95\hsize}{!}{$
	\Omega_0 = \left\{\vect{X} = [R \cos\Theta,\,R\sin\Theta,Z]\in\R^3 \;
	|\;R_\text{i}\leq R<R_\text{o}\text{ and }0\leq\Theta<2\pi\text{ and }0<Z<L\right\},
	$}
\]
be the reference configuration of the HT, where $R_\text{i}$ and $R_\text{o}$ are respectively the internal and the external radius of the cylinder, $L$ is the height of the hollow cylinder and $\vect{X}$ is the material position coordinate. Although the HT is composed by different layers, for sake of simplicity, we assume that the HT is a homogeneous one-layered tissue. By experimental evidences, the HT can be modeled as a hollow cylinder since there is a small lumen in the middle. 
We indicate with $\vect{\varphi}:\Omega_0\rightarrow\R^3$ the deformation field, 
so that the actual configuration of the body $\Omega$ is given by $\vect{\varphi}(\Omega_0)$. 
Let $\vect{x}= \vect{\varphi}(\vect{X})$ be the actual position and the displacement vector is defined as 
$\vect{u}(\vect{X}) = \vect{\varphi}(\vect{X})-\vect{X}$. 
Let $\tens{F}$ be the deformation gradient, i.e. $\tens{F} = \Grad\vect{\varphi}$.

In order to describe the finite torsion induced by tissue remodelling, we consider the multiplicative decomposition of the deformation gradient \cite{kroner1959allgemeine,lee1969elastic,rodriguez1994stress}, such as
\begin{equation}
\label{eq:dec_molt}
\tens{F} = \tens{F}_{\rm e} \tens{G},
\end{equation}
where $\tens{G}$ describes the metric distortion induced by the applied torsion and $\tens{F}_{\rm e}$ is the elastic deformation of the material restoring the geometrical compatibility of the current configuration. Defining $\gamma$ as the finite torsion rate resulting from the remodelling processes, we choose $\tens{G}$ as
\begin{equation}
\label{eq:tens_G}
\tens{G} = \begin{pmatrix}
1&0&0\\
0&1&-\gamma R\\
0&0&1
\end{pmatrix}.
\end{equation}
We further assume that the cylinder cannot elongate along the $z$-direction, {\em i.e.} 
\begin{equation}
\label{eq:zL}
z(Z = \pm L) = \pm L.
\end{equation}
Since no relevant growth processes occur in the stages of interest of c-looping (${\rm det}\, \tens{G}=1$) and the tissue is mainly composed by water, we model the HT as incompressible media, namely we enforce that
\begin{equation}
\label{eq:incompressibility}
{\rm det}\, \tens{F}_{\rm e} = 1.
\end{equation}

%We now introduce some mechanical constitutive assumptions.

\subsection{Boundary-Value Problem (BVP)}
\label{subsec:constitutive_assumptions}
Since we are describing the first stage of development and fibers are not yet present \cite{hamburger1951series}, it is reasonable to model the HT as an isotropic body. We assume that the HT is composed of a homogeneous hyperelastic material, 
having strain energy density $\psi$. 
The first Piola-Kirchhoff stress tensor $\tens{P}$  and the Cauchy stress tensors $\tens{T}$ are then given by
\[
\tens{P} = {\rm det}\,\tens{G}\frac{\partial\psi(\tens{F\tens{G}^{-1}})}{\partial\tens{F}}-p\tens{F}^{-1}\qquad\tens{T} = \frac{1}{{\rm det}\,\tens{F}}\tens{F}\tens{P}
\] 
where $p$ is the Lagrange multiplier enforcing the incompressibility constraint, i.e. ${\rm det}\, \tens{F}_{\rm e}= 1$. Assuming quasi-static conditions in absence of external body forces, the balance of the linear momentum  reads
\begin{equation}
\label{eq:balance}
\Diver\tens{P} = \vect{0}\text{ in }\Omega_{0},\qquad\text{or}\qquad\diver\tens{T}=\vect{0}\text{ in }\Omega
\end{equation}
where $\Diver$ and $\diver$ denote the divergence operator in material and current frame, respectively.
The nonlinear system of equations \eqref{eq:balance} is complemented by the following Neumann condition on the inner and outer boundaries
\begin{equation}
\label{eq:noloads}
\left\{
\begin{aligned}
\tens{T}\cdot \vect{n} &= \vect{0} \quad \hbox{on } r = r_\text{i}\\
\tens{T}\cdot \vect{n} &= \vect{0} \quad \hbox{on } r = r_\text{o}   
\end{aligned}
\right. , 
\end{equation}
where $\vect{n}$ is the outer normal in spatial coordinates, and $r_\text{i}, r_\text{o}$ are the spatial inner and outer radius, respectively. By performing a pull-back of Eq. \eqref{eq:noloads}, the Lagrangian form of the boundary condition is given by
\begin{equation}
\label{eq:noloadsP}
\left\{
\begin{aligned}
\tens{P}^T\cdot \vect{N} &= \vect{0} \quad \hbox{on } R = R_\text{i}\\
\tens{P}^T\cdot \vect{N} &= \vect{0} \quad \hbox{on } R = R_\text{o}   
\end{aligned}
\right. 
\end{equation}
where $\vect{N}$ is the material outer normal. To keep the model as simple as possible, we 
assume that the tube behaves as a neo-Hookean material with the strain energy density given by
\begin{equation}
\label{eq:strain_energy}
W(\tens{F}) =\det(\tens{G})W_0(\tens{F}_\text{e}) =(\det\tens{G})\frac{\mu}{2}\left(\tr(\tens{F}_\text{e}^T\tens{F}_\text{e})-3\right).
%\psi(\tens{F}) =\frac{\mu}{2}\left(\tr(\tens{F}^T\tens{F})-3\right).
\end{equation}
The first Piola--Kirchhoff and Cauchy stress tensors read respectively
\begin{equation}
\label{eq:stress_tensors}
\left\{
\begin{aligned}
&\tens{P} = \mu  ({\rm det}\, \tens{G}) \tens{G}^{-1}\tens{G}^{-T}\tens{F}^{T} - p\tens{F}^{-1},\\
&\tens{T} = \mu \tens{F}\tens{G}^{-1}\tens{G}^{-T}\tens{F}^{T} - p\tens{I}.
\end{aligned}
\right.
\end{equation}
%where $\tens{B}= \tens{F}\tens{F}^{T}$ is the left Cauchy-Green tensor.
Eqs.~\eqref{eq:incompressibility}, \eqref{eq:balance} and \eqref{eq:noloadsP} define the nonlinear elastic BVP. 
\subsection{Radially-symmetric solution}
\label{subsec:base_solution}
Let $(\vect{E}_R,\,\vect{E}_\Theta,\,\vect{E}_Z)$ and $(\vect{e}_r,\,\vect{e}_\theta,\,\vect{e}_z)$ 
be the unit vectors in material and spatial polar coordinates, respectively.
Denoting  $(r,\,\theta,\,z)$ the polar coordinates of a point, we search for a radially-symmetric solution 
\[
\vect{\varphi}(\vect{X}) = r(R)\vect{e}_r + Z\vect{e}_z.
\]
so that  the geometrical and the elastic  deformation gradient read 
\begin{equation}
\label{eq:deformationgradient}
\tens{F}=
{\rm diag} \left(\frac{\partial r(R)}{\partial R},\, \frac{r}{R},\,1\right), \qquad \tens{F}_{\rm e} = \tens{F}\tens{G}^{-1}=\begin{pmatrix}
\frac{\partial r(R)}{\partial R}&0&0\\
0&\frac{r}{R}&\gamma R\\
0&0&1
\end{pmatrix}.
\end{equation}

From  Eq. \eqref{eq:deformationgradient} and  the incompressibility
constraint Eq. \eqref{eq:incompressibility}, we get $r'r = R$,
where $'$ denotes differentiation. By integrating, we obtain
\begin{equation}
\label{eq:r}
r(R) = \sqrt{R^2 +r_\text{i}^2-R_\text{i}^2}.
\end{equation}
The balance of the linear  momentum in polar coordinates imposes
\begin{equation}
\label{eq:divTrad}
\frac{d T_{rr}}{dr} + \frac{T_{rr}-T_{\theta\theta}}{r}=0
\end{equation}
where $T_{hk}$, with $h,k$ spanning over $(r,\theta,z)$, are the components of the Cauchy stress tensor $\tens{T}$ in polar coordinates. Using Eqs. \eqref{eq:deformationgradient} and \eqref{eq:stress_tensors}, the Cauchy stress tensor is given by
\begin{equation}
\label{eq:comp_tensT}
\tens{T}=\left(
\begin{array}{ccc}
\frac{\mu  \left(r^2-r_{\rm i}^2+R_{\rm i}^2\right)}{r^2}-p & 0 & 0 \\
0 & \mu  r^2 \left(\gamma ^2+\frac{1}{r^2-r_{\rm i}^2+R_{\rm i}^2}\right)-p & \gamma  \mu  r \\
0 & \gamma  \mu  r & \mu -p \\
\end{array}
\right).
\end{equation}
As regarding the current radii, by using Eq. \eqref{eq:r}, the expression of the external radius $r_{\rm o}$ is given by
\begin{equation}
\label{eq:r_o}
r_{\rm o} = \sqrt{R_{\rm o}^2 +r_{\rm i}^2 - R_{\rm i}^2},
\end{equation}
while for the internal one, we have to integrate Eq. \eqref{eq:divTrad} from $r_{\rm o}$ to $r_{\rm i}$, use the boundary conditions Eq. \eqref{eq:noloads} to get
\begin{equation}
\label{eq:rel_ro_ri}
\begin{footnotesize}
\begin{aligned}
\frac{1}{2} \mu  \left(\left(R_{\rm i}^2-R_{\rm o}^2\right) \left(\gamma ^2+\frac{r_{\rm i}^2-R_{\rm i}^2}{r_{\rm i}^2
	\left(r_{\rm i}^2-R_{\rm i}^2+R_{\rm o}^2\right)}\right)+\log
\left(\frac{r_{\rm i}^2-R_{\rm i}^2+R_{\rm o}^2}{R_{\rm o}^2}\right)+\log
\left(\frac{R_{\rm i}^2}{r_{\rm i}^2}\right)\right)=0
\end{aligned}
\end{footnotesize}
\end{equation}
which is an implicit relation to derive first $r_{\rm i}$ and then $r_{\rm o}$ from Eq. \eqref{eq:r_o}. It can be solved by using a Newton method fixing $\gamma$ and the initial geometry of the hollow tube.

Finally, for the Lagrange multiplier $p$, we can integrate Eq. \eqref{eq:divTrad} from $r$ to $r_{\rm o}$, obtaining
\begin{equation}
\label{eq:Trr}
T_{rr}(r) = \int^{r_\text{o}}_r\left[\frac{\mu  \left(\rho ^2-r_{\rm i}^2+R_{\rm i}^2\right)}{\rho ^3}-\mu  \rho  \left(\gamma ^2+\frac{1}{\rho
	^2-r_{\rm i}^2+R_{\rm i}^2}\right)\right]\,d\rho.
\end{equation}
We can find the pressure field by solving Eq. \eqref{eq:Trr} with respect to $p$, which reads
\begin{equation}
\label{eq:p}
\begin{footnotesize}
\begin{multlined}
p(r) = \frac{1}{2} \mu  \left(\gamma ^2
\left(-r^2+r_{\rm i}^2-R_{\rm i}^2+R_{\rm o}^2\right)+\frac{R_{\rm i}^2-r_{\rm i}^2}{r^2}+\frac{R_{\rm i}^2-r_{\rm i}^2}{r_{\rm i}^2-R_{\rm i}^2+R_{\rm o}^2}\right.\\
\left.+\log
\left(\frac{r^2}{r^2-r_{\rm i}^2+R_{\rm i}^2}\right)+\log \left(\frac{R_{\rm o}^2}{r_{\rm i}^2-R_{\rm i}^2+R_{\rm o}^2}\right)+2\right)
\end{multlined}
\end{footnotesize}
\end{equation}
Eqs. \eqref{eq:r} and \eqref{eq:p} represent the radially symmetric solution of the BVP.

\section{Linear stability analysis}
\label{sec:lin_stab_clooping}
In this section we study the linear stability of the finitely deformed tube by using the method of incremental deformations superposed on a finite strain \cite{ogden1997non}.
We rewrite the resulting incremental BVP into the Stroh formulation that is solved using the impedance matrix method.

\subsection{Incremental BVP}
\label{subsec:incr_eq}
We apply the theory of incremental deformations superposed on finite strains  to investigate the stability of the radially symmetric solution.  Let $\delta \vect{u}$ be the incremental displacement field and let $\tens{\Gamma} = \grad\delta \vect{u}$.  We introduce the push-forward of the incremental Piola-Kirchhoff  stress tensor $\delta \tens{P}_0$
in the axis-symmetric deformed configuration, given by
\begin{equation}
\label{eq:A0}
\delta \tens{P} = \mathcal{A}_0 : \tens{\Gamma} +p \tens{\Gamma} - \delta p \tens{I},
\end{equation}
where $\mathcal{A}_0$ is the fourth order tensor of instantaneous elastic moduli, $\delta p$ is the increment of the Lagrangian multiplier that imposes the incompressibility constraint. The two dots operator $(:)$ denotes the double contraction of the indices
\[
(\mathcal{A}_0 : \tens{\Gamma})_{rs} = (A_0)_{rshk}\Gamma_{kh},
\]
where the convention of summation over repeated indices is used. The components of the tensor $\mathcal{A}_0$ for a neo-Hookean material are given by 
\[
(A_0)_{rshk} = \mu \delta_{rk}(B_\text{e})_{sh} 
\]
where $\tens{B}_\text{e} = \tens{F}_\text{e} \tens{F}_\text{e}^T$ and $\delta_{rk}$ is the Kronecker delta. 
The incremental equilibrium equation and the linearised form of the incompressibility constraint read respectively
\begin{equation}
\left\{
\begin{aligned}
\label{eq:deltaP}
&\Diver\delta\tens{P} = \vect{0} &&\hbox{in } \Omega, \\
&\tr \, \tens{\Gamma} = 0 &&\hbox{in } \Omega.
\end{aligned}
\right.
\end{equation}
This system of partial differential equations is complemented by the following boundary conditions
\begin{equation}
\left\{
\begin{aligned}
\label{eq:bcincremented}
&\delta\tens{P}^{T} \,\vect{e}_{r}= 0 &&\hbox{on } r = r_{\rm i}, \\
&\delta\tens{P}^{T} \,\vect{e}_{r} = 0 &&\hbox{on } r = r_{\rm o}.
%&\delta\tens{P}^{T} \,\vect{e}_{r} \cdot \vect{e}_{r} = 0&&\hbox{on } z = 0,  z = L,\\
%&\delta\tens{P}^{T} \,\vect{e}_{r} \cdot \vect{e}_{\vartheta} = 0&&\hbox{on } z = 0,  z = L,\\
%&\delta u_{z} = 0 &&\hbox{on } z = 0,  z = L.
\end{aligned}
\right.
\end{equation}
%and $L$ is the length of the cylinder.
\subsection{Stroh formulation}
\label{subsec:Stroh}
%We rewrite the incremental problem in a non-dimensional form adopting as the characteristic length 
%scale $R_{\rm o}$. The behaviour of the problem is governed by the non-dimensional parameter
%\begin{equation}
%\label{eq:par_adim}
%\alpha_L = \frac{L}{R_{\rm o}},
%\end{equation}
%in addition to the torsion parameter $\gamma$.
We denote with $u$, $v$ and $w$ the components of $\delta u$ in cylindrical coordinates and with $\delta P_{rr}$, $\delta P_{r\vartheta}$ and $\delta P_{rz}$ the components of the incremental stress tensor. Following \cite{Stroh_1962}, we rewrite the system of partial differential equations Eq. \eqref{eq:deltaP} into a system of ordinary differential equations, by assuming the following variable separation ansatz for the incremental  fields \cite{Ciarletta_2014,balbi2015helical}
\begin{align}
\label{eq:uincr}
u(r,\vartheta,z)&=U(r) \cos(k z - m\vartheta),\\
\label{eq:vincr}
v(r,\vartheta,z)&=V(r) \sin(k z - m\vartheta)\\
\label{eq:wincr}
w(r,\vartheta,z)&=W(r) \sin(k z - m\vartheta),\\
\label{eq:prr_incr}
\delta P_{rr}(r,\vartheta,z)&= s_{rr}(r) \cos(k z - m\vartheta)\\
\label{eq:prtheta_incr}
\delta P_{r\vartheta}(r,\vartheta,z) &= s_{r\vartheta}(r) \sin(k z - m\vartheta),\\
\label{eq:prz_incr}
\delta P_{rz}(r,\vartheta,z) &= s_{rz}(r) \sin(k z - m\vartheta),\\
\label{eq:deltapres}
\delta p(r,\vartheta,z) &= Q(r)\cos(k z - m\vartheta),
\end{align}
where $m$ and  $k=(2n\pi)/L$ are respectively the circumferential and axial wavenumbers, with $m,n\in\N$.  By substituting Eq. \eqref{eq:prr_incr} into Eq. \eqref{eq:A0}, we get
$$ Q(r) = \frac{U'(r) \left(r^2 p(r)+r^2-r_{\rm i}^2+R_{\rm i}^2\right)}{r^2}-s_{rr}(r), $$
where $p(r)$ is defined in Eq.~\eqref{eq:p}. Following a similar and well established procedure \cite{balbi2015helical}, the incremental problem can be rewritten into the Stroh form \cite{Stroh_1962}, such as
\begin{equation}
\label{eq:stroh}
\frac{d \vect{\eta}}{d r} = \frac{1}{r} \tens{N} \vect{\eta},
\end{equation}
where $\vect{\eta}$ is the \emph{displacement-traction vector} defined as
\[
\vect{\eta} = [\vect{U}, r\vect{\Sigma}] \quad \hbox{where} \quad \left\{
\begin{aligned}
&\vect{U} = [U,V,W],\\
&\vect{\Sigma}= [s_{rr}, s_{r\vartheta},s_{rz}].
\end{aligned}
\right.
\]
The matrix $\tens{N} \in \mathbb{R}^{6\times6}$ is the \emph{Stroh matrix} and it has the following sub-block form
\[
\tens{N}=
\begin{bmatrix}
\tens{N}_1 &\tens{N}_2\\
\tens{N}_3 &\tens{N}_4
\end{bmatrix},
\]
such that $\tens{N}_{i} \in \R^3 \times \R^3$, $\tens{N}_1 = - \tens{N}_4^T$, $\tens{N}_2 = \tens{N}_2^T$ and $\tens{N}_3 = \tens{N}_3^T$, which means that the Stroh matrix has an Hamiltonian structure \cite{Fu_2007}. The expression of the four blocks is given by
\begin{equation}
\label{eq:block_stroh}
\begin{aligned}
&\tens{N}_1=
\begin{bmatrix}
-1 & m & -k r \\
-m \sigma p & \sigma p& 0 \\
k r p \sigma & 0 & 0 \\
\end{bmatrix},
&&\tens{N}_2 = 
\begin{bmatrix}  
0 & 0 & 0 \\
0 & \sigma & 0 \\
0 & 0 & \sigma \\
\end{bmatrix},
&\tens{N}_3= 
\begin{bmatrix}  
\kappa_{11}& \kappa_{12}& \kappa_{13} \\
\kappa_{12} &\kappa_{22}& \kappa_{23} \\
\kappa_{13} & \kappa_{23} & \kappa_{33}
\end{bmatrix},
\end{aligned}
\end{equation}
where 
\begin{equation*}
\resizebox{0.95\textwidth}{!}{$ \begin{split}
	&\kappa_{11} = 1/\sigma + r^2 \delta_1 + 2 p + \delta_3 \sigma &&\kappa_{12} = \sigma m (1/\sigma^2 -2 + p^2)+ \delta_2\\
	&\kappa_{13} =k/r (\sigma + r p) &&\kappa_{22} = r^2 (k + \gamma^2 - 2m\gamma k) + \sigma (1-p^2) + m^2(2 + \sigma \delta_5 + \sigma \gamma R)\\
	&\kappa_{23} = -k m (\sigma/r + 2 p r)
	&&\kappa_{33} = m^2(\sigma + r \gamma^2) + r^2 k \delta_4 
	\end{split}$}
\end{equation*}
and
\begin{equation*}
\resizebox{0.95\textwidth}{!}{$ \begin{split}
	&\sigma = \frac{r^2}{R}&&\delta_1 = \gamma ^2+k^2-2 \gamma  k m+\gamma ^2 m^2 &&\delta_2 = -p^2 \left(k^2 r^2+m^2\right)+m^2+1\\
	&\delta_3 = 2 r^2 \gamma (k-m \gamma) -2 m p &&\delta_4 = k(1 + 1/\sigma - 2p) + 2 m \gamma &&\delta_5 = (R_{\rm i}^2 -r_{\rm i}^2)^2/r^4,
	\end{split}$}
\end{equation*}
with $r$ and $p$ defined in Eq. \eqref{eq:r} and Eq. \eqref{eq:p}, respectively.
Eq. \eqref{eq:stroh} with the boundary condition $\vect{\Sigma} = \vect{0}$ at $r = r_{\rm i}$ and $r = r_{\rm o}$ define the incremental BVP.

\subsection{Impedence matrix method}
\label{subsec:impedence}
The incremental BVP is numerically solved using the impedance matrix method \cite{biryukov1985impedance,Biryukov_1995}. 
Following a similar procedure used in \cite{balbi2015helical}, we introduce the matricant
\[
\tens{M}(r,r_{\text o}) = \begin{bmatrix}
\tens{M}_1(r,r_{\text o}) &\tens{M}_2(r,r_{\text o})\\
\tens{M}_3(r,r_{\text o})&\tens{M}_4(r,r_{\text o})
\end{bmatrix}, \qquad \tens{M}(r,r_{\text o}) \in \R^{6\times6},
\]
defined as the solution of the problem
\begin{equation}
\label{eq:matricant}
\frac{d\tens{M}}{dr} = \frac{1}{r}\tens{N}\tens{M}(r,r_{\text o}),\quad \tens{M}(r_{\text o},r_{\text o}) = \tens{I}.
\end{equation}
Since the solution of the Stroh problem Eq. \eqref{eq:stroh} can be expressed as
$$
\vect{\eta}(r) = \tens{M}(r,r_{\rm o}) \vect{\eta}(r_{\rm i})
$$
and no traction loads are applied on the external surface, i.e. $
\vect{\Sigma}(r_{\rm o}) = \vect{0}$, we can define the {\em conditional impedence matrix} $\tens{Z}(r, r_{\rm o})$ \cite{Norris_2010} as
\begin{equation}
\label{eq:Z}
\tens{Z}(r,r_{\rm o})=\tens{M}_3(r,r_{\rm o})\tens{M}^{-1}_3(r,r_{\rm o}),
\end{equation}
where the term {\em conditional} refers to the dependence on the boundary condition at $r_{\rm o}$. Omitting the explicit dependence of $Z$ on $r$ and $r_{\rm o}$, such
a matrix satisfy the following relation
\begin{equation}
\label{eq:imp_matrix}
r\vect{\Sigma} = \tens{Z}\vect{U} \qquad \forall r \in (r_{\rm i},r_{\rm o})
\end{equation}
By using Eq.~\eqref{eq:imp_matrix}, we can rewrite the Stroh problem given by Eq.~\eqref{eq:stroh} into a differential Riccati equation, such as
\begin{gather}
\label{eq:riccatiU}
\frac{d \vect{U}}{d r} = \frac{1}{r} \left(\tens{N}_1+\tens{N}_2\tens{Z}\right)\vect{U},\\
\label{eq:riccatiZU}
\frac{d\tens{Z}}{dr}\vect{U}+\tens{Z}\frac{d\vect{U}}{dr} = \frac{1}{r} \left(\tens{N}_3+\tens{N}_4\tens{Z}\right)\vect{U}.
\end{gather}
Substituting Eq.~\eqref{eq:riccatiU} into Eq.~\eqref{eq:riccatiZU} we get the following differential Riccati equation
\begin{equation}
\label{eq:riccati}
\frac{d \tens{Z}}{d r} = \frac{1}{r} \left(\tens{N}_3+\tens{N}_4\tens{Z}-\tens{Z}\tens{N}_1-\tens{Z}\tens{N}_2\tens{Z}\right).
\end{equation}
We integrate Eq.~\eqref{eq:riccati} from $r_{\text o}$ to $r_{\text i}$, using as starting condition the fact that there are no applied loads at $r = r_{\rm o}$, i.e.
\[
\tens{Z}(r_\text{o},\,r_\text{o}) = \vect{0}.
\]
To construct a bifurcation criterion, we use the fact that there are no applied loads in $r= r_{\rm i}$: non-null solutions of the incremental problem exist if and only if
\begin{equation}
\label{eq:stopcondition}
\det \tens{Z}(r_{\rm i})=0.
\end{equation}
Fixing the initial geometry of the HT, making outer iterations on the wavenumbers $m$ and $k$, for a fixed value of the control parameter $\gamma$ we integrate the Riccati Eq.~\eqref{eq:riccati} from $r= r_{\text o}$ up to $r = r_{\text i}$ making use of the the software \textsc{Mathematica} 11.3 (Wolfram Research, Champaign, IL, USA). We iteratively increase the control parameter $\gamma$ following its stable solution \cite{Ciarletta_2014} until the bifurcation criterion Eq.~\eqref{eq:stopcondition} is satisfied.

\subsection{Marginal stability thresholds}
\label{subsec:resultt_linstab}
In this section we collect and discuss the results of the linear stability analysis. First of all, we introduce the dimensionless parameters which govern the boundary value problem, i.e.
\begin{align}
\label{eq:dim_parameters}
&\tilde{k} = k R_{\rm o}&&\alpha_R =  \frac{R_{\rm o}}{R_{\rm i}}&&\alpha_L = \frac{L}{R_{\rm o}},
\end{align}
where $\tilde{k}$ represents the dimensionless axial wavenumber, $\alpha_R$  and $\alpha_L $  are geometrical parameters, representing  the thickness and the slenderness ratios, respectively. Given a material  length $L$, the admissible axial wavenumber $\tilde{k}$ is given by
\begin{align}
\label{eq:k_discreto}
\tilde{k} = \frac{2 \pi n}{\alpha_L} \qquad n \in \N.
\end{align}
Here, we aim at characterizing the torsion deformation during the c-looping just by considering suitable biological values of the two parameters involved, the initial aspect ratio $\alpha_R$ of the HT and its length $\alpha_L$. A lot of experiments have been performed on animals whose cardiovascular system is similar to humans, for instance the chick embryo \cite{voronov2004role,Zamir_2004, ramasubramanian2006computational, manner2014cardiac, shi2014bending}. From these experimental results,  a physiologically relevant range for both the initial aspect ratio $\alpha_R$ and the dimensionless length $\alpha_L$ is
\begin{align}
\label{eq:int_adim}
&\alpha_R = [1.35, 2.85] &&\alpha_L = [7,10].
\end{align}

\begin{figure}[h!]
	\begin{subfigure}{.3\linewidth}
		\centering
		\includegraphics[width=\textwidth]{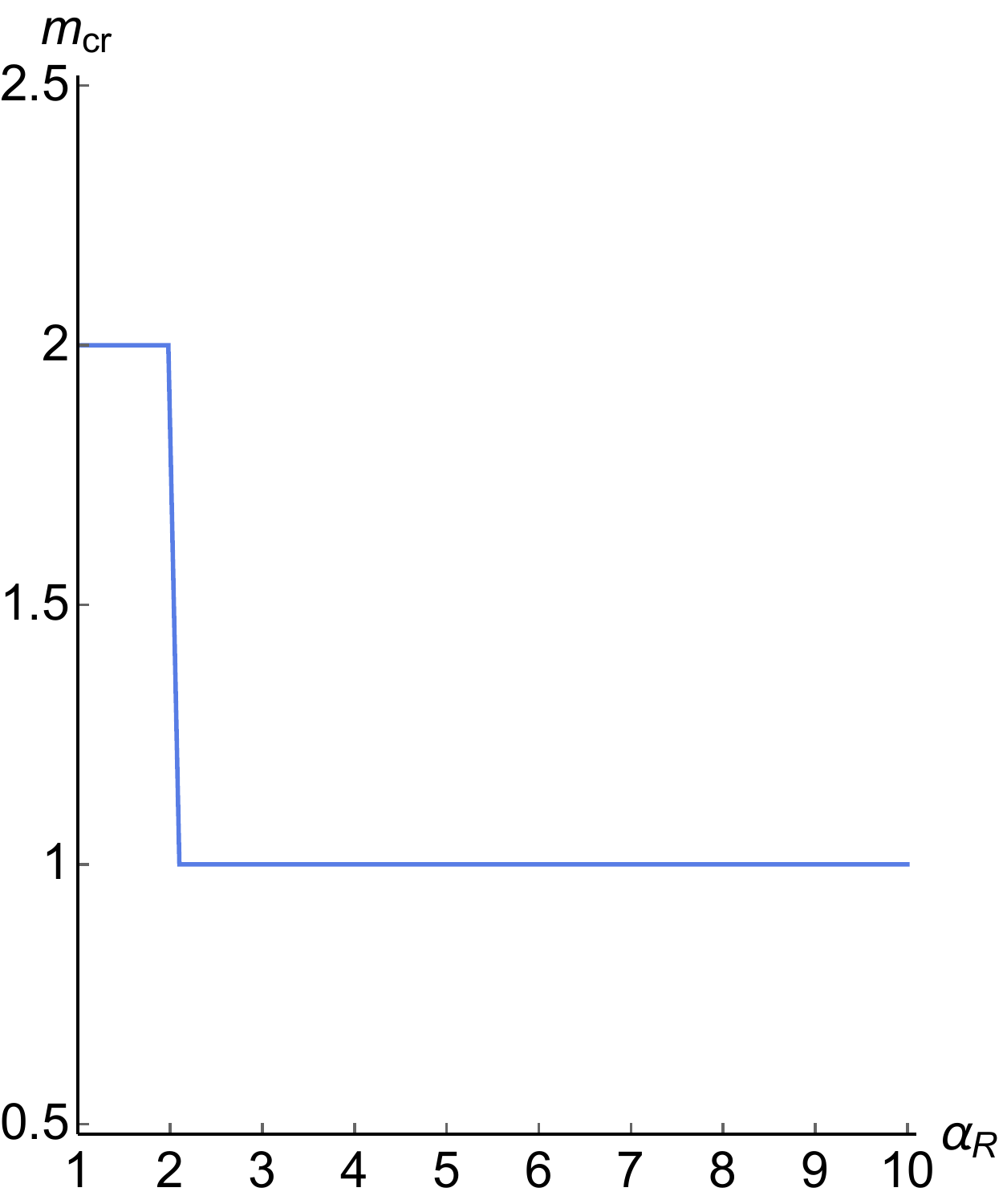}
		\caption{$\alpha_R \in (1, 10)$}
		\label{fig:mvsalphaR}
	\end{subfigure}
	\begin{subfigure}{.3\linewidth}
		\centering
		\includegraphics[width=0.84\textwidth]{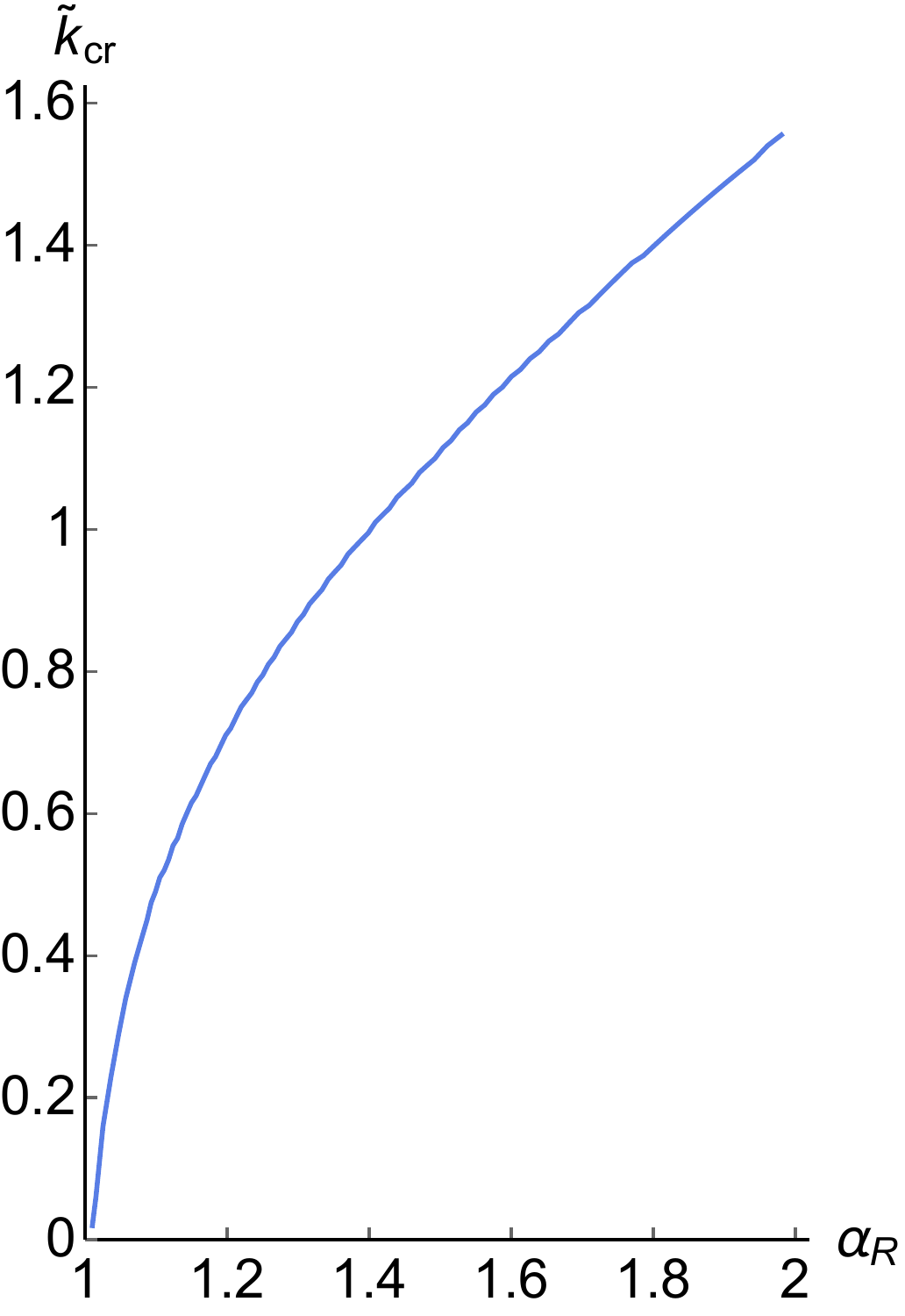}
		\caption{$\alpha_R \in (2,10)$}
		\label{fig:kzvsalphaR_mcr2}
	\end{subfigure}
	\begin{subfigure}{.3\linewidth}
		\centering
		\includegraphics[width=\textwidth]{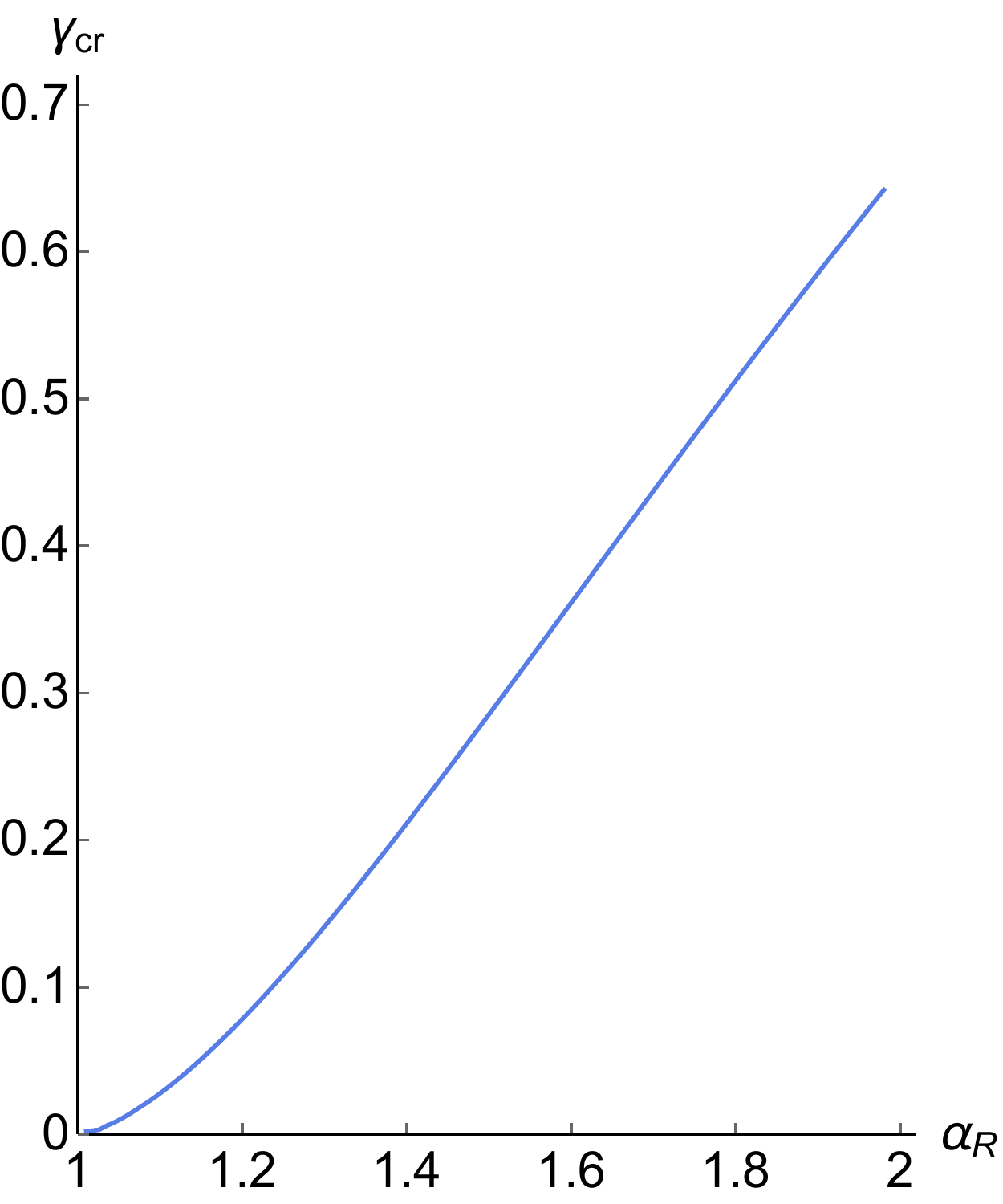}
		\caption{$\alpha_R \in (1, 2)$}
		\label{fig:gammavsalphaR_mcr2}
	\end{subfigure}\\
	\begin{subfigure}{.44\linewidth}
		\centering
		\includegraphics[width=0.65\textwidth]{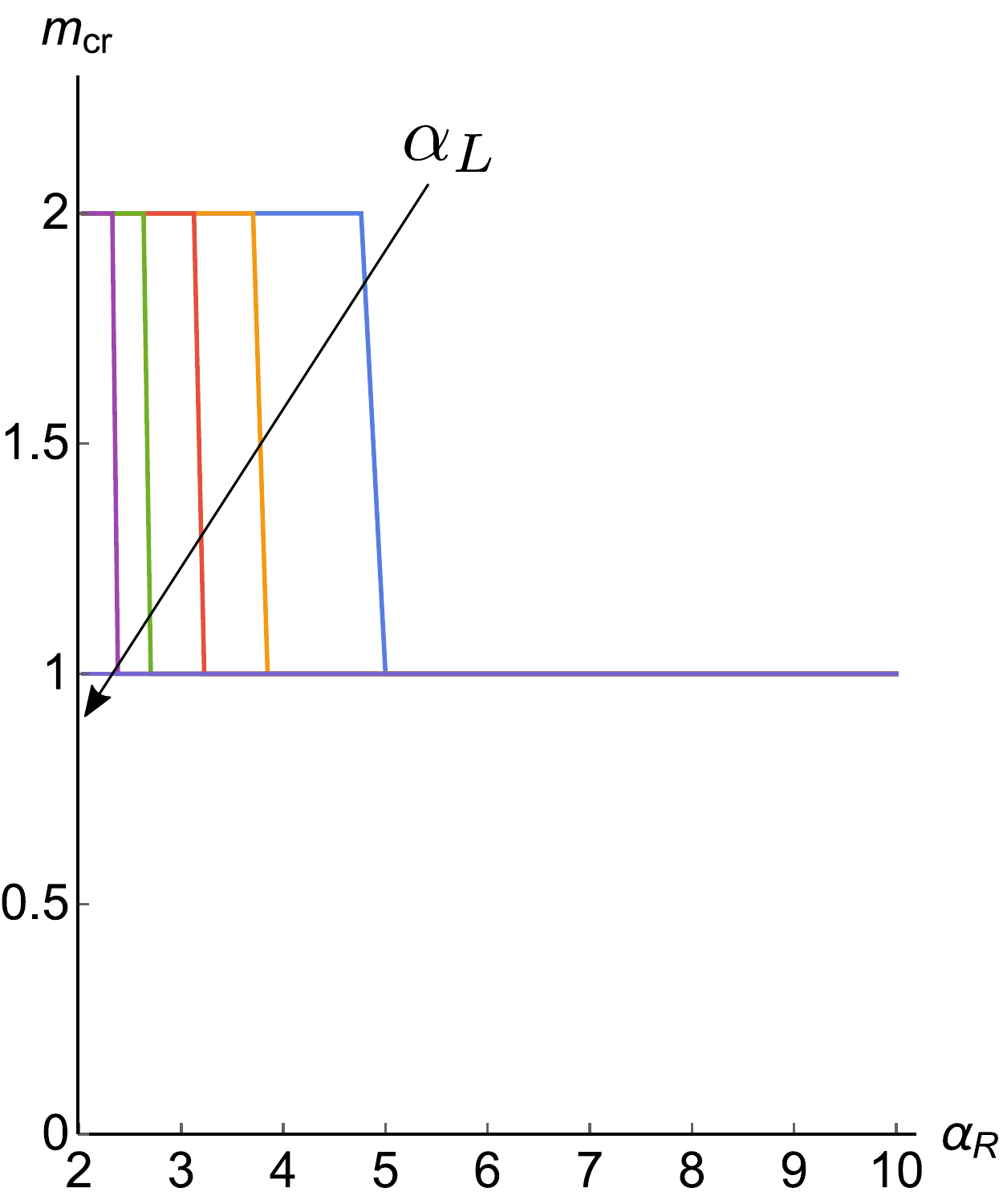}
		\caption{$\alpha_R \in (2, 10)$}
		\label{fig:mvsalphaR_varioalphaL}
	\end{subfigure}
	\begin{subfigure}{.44\linewidth}
		\centering
		\includegraphics[width=0.65\textwidth]{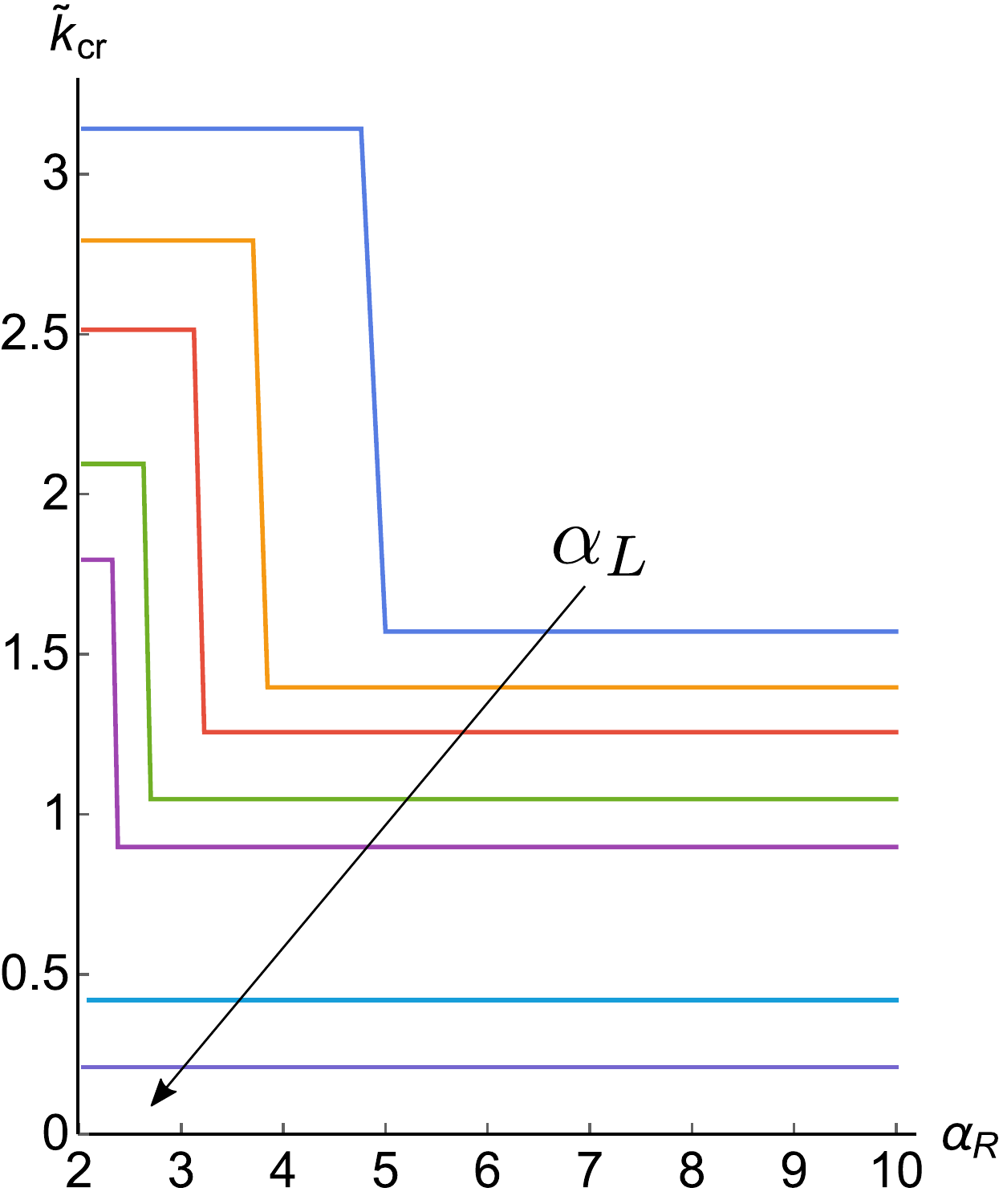}
		\caption{$\alpha_R \in (2, 10)$}
		\label{fig:kzvsalphaR_varioalphaL}
	\end{subfigure}
	\caption{Plot of the critical values versus $\alpha_R$, varying the initial aspect $\alpha_R \in (1,10]$. (a) Plot of the circumferential critical wavenumber $m_{\rm cr}$ versus $\alpha_R$. Varying $\alpha_R \in (1,2)$ and $\alpha_L = + \infty$, (b) plot of the critical axial wavenumber $\tilde{k}_{\rm cr}$ and (c) of the marginal stability threshold $\gamma_{\rm cr}$ versus $\alpha_R$. Varying $\alpha_R \in (2,10)$, (d)  plot of the critical circumferential wavenumber $m_{\rm cr}$ and (e) of the critical axial wavenumber $\tilde{k}_{\rm cr}$ versus $\alpha_R$ having chosen different values of $\alpha_L = \{4,4.5,5,6,7,15,30\}$.}
	\label{fig:vsalphaR}
\end{figure}

In the following, for sake of simplicity, we  present the results  considering $\tilde{k}$ as a continuous quantity and varying $\alpha_R \in (1, + \infty)$, since  neither $\alpha_R = 1$ nor $\alpha_R \to +\infty$ represent a hollow cylinder. In Fig. \ref{fig:vsalphaR}, we plot the marginal stability threshold $\gamma_{\rm cr}$, the circumferential $m_{\rm cr}$ and the axial $\tilde{k}_{\rm cr}$ critical wavenumbers versus the initial aspect ratio $\alpha_R$. For graphical clarity, we split the two cases: $\alpha_R \in (1,2)$, in which the critical circumferential number $m_{\rm cr} = 2$ and $\alpha_R \in (2, +\infty)$, in which the critical circumferential number depends on the length of the tube, i.e. on $\alpha_L$. 

Considering the infinite cylinder, i.e. $\tilde{k} \in \R$, in Figs. \ref{fig:mvsalphaR}, we plot the critical circumferential number versus $\alpha_R$ and we notice that there is a cut-off thickness at which the morphological transition of the HT completely changes: for thin cylinder $m_{\rm cr} = 2$, while for $\alpha_R > 2$, the critical circumferential number is $m_{\rm cr}=1$. Fixing $\alpha_R \in (1,2)$, we plot the critical axial wavenumber $\tilde{k}_{\rm cr}$ and the marginal stability threshold $\gamma_{\rm cr}$, see respectively Figs. \ref{fig:kzvsalphaR_mcr2} - \ref{fig:gammavsalphaR_mcr2}. Both the marginal stability threshold $\gamma_{\rm cr}$ and the critical axial wavenumber increase as $\alpha_R$ increases: as already pointed out in \cite{balbi2015helical}, a thin hollow cylinder buckles earlier then a thick one (compare their Figs. 7a - 8a with our Figs. \ref{fig:gammavsalphaR_mcr2} - \ref{fig:kzvsalphaR_mcr2}). 

\begin{figure}
	\centering
	\includegraphics[width=0.6\textwidth]{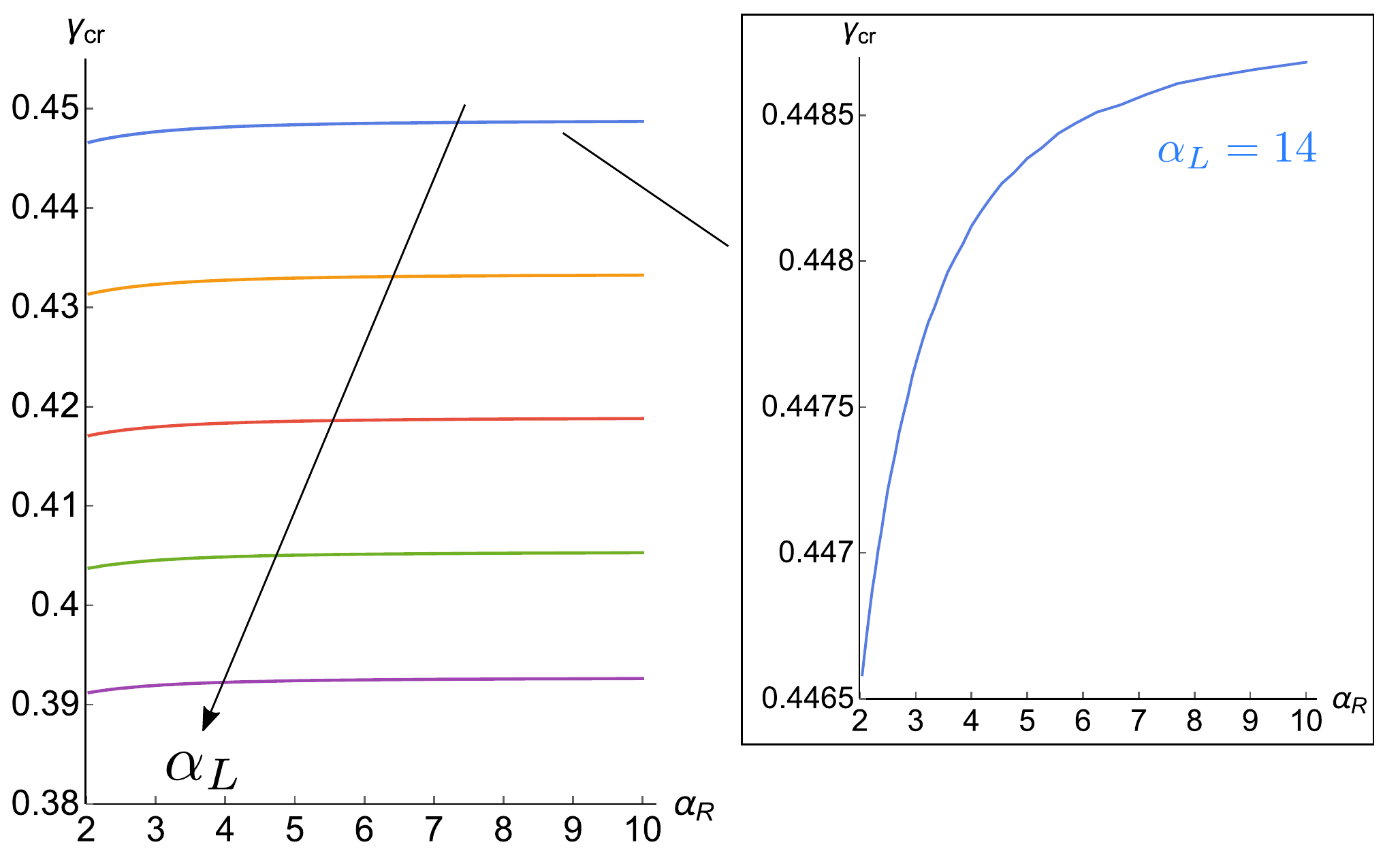}
	\caption{Plot of the marginal stability threshold $\gamma_{\rm cr}$ versus $\alpha_R$ varying $\alpha_L= \{14,14.5,15,15.5,16\}$. The inset shows zoom of the curve with  $\alpha_L = 14$.}
	\label{fig:gammavsalphaR_mcr1}
\end{figure}

Since an infinite thick cylinder immediately buckles for a very small value of the control parameter $\gamma$, we consider finite hollow thick tubes varying $\alpha_R \in (2,10)$ and \\$\alpha_L = \{4, 4.5 ,5, 6, 7, 15, 30\}$. From Fig. \ref{fig:mvsalphaR_varioalphaL} and Fig. \ref{fig:kzvsalphaR_varioalphaL}, we notice that both the critical wavenumbers $m_{\rm cr}$ and $\tilde{k}_{\rm cr}$ are affected by the length of the cylinder. For a short and thin cylinder, the critical circumferential wavenumber $m_{\rm cr} = 2$ and there is a doubling of axial wavenumber, i.e. $\tilde{k}_{\rm cr} = 4 \pi /\alpha_L$. As the cylinder gets thicker, we find that $m_{\rm cr} = 1$ and $\tilde{k}_{\rm cr} = 2\pi/\alpha_L$, similar to what happens for an infinite one. Finally, in Fig. \ref{fig:gammavsalphaR_mcr1}, we plot the marginal stability threshold $\gamma_{\rm cr}$ versus $\alpha_R$ for different values of $\alpha_L$. For graphical scaling of the parameters, we change the range of $\alpha_L = \{14,14.5,15,15.5,16\}$: similar to what happens for thin cylinder, $\gamma_{\rm cr}$ increases as $\alpha_R$ increases, see the inset in Fig. \ref{fig:gammavsalphaR_mcr1}.

\begin{figure}[h!]
	\begin{subfigure}{.32\linewidth}
		\centering
		\includegraphics[width=\textwidth]{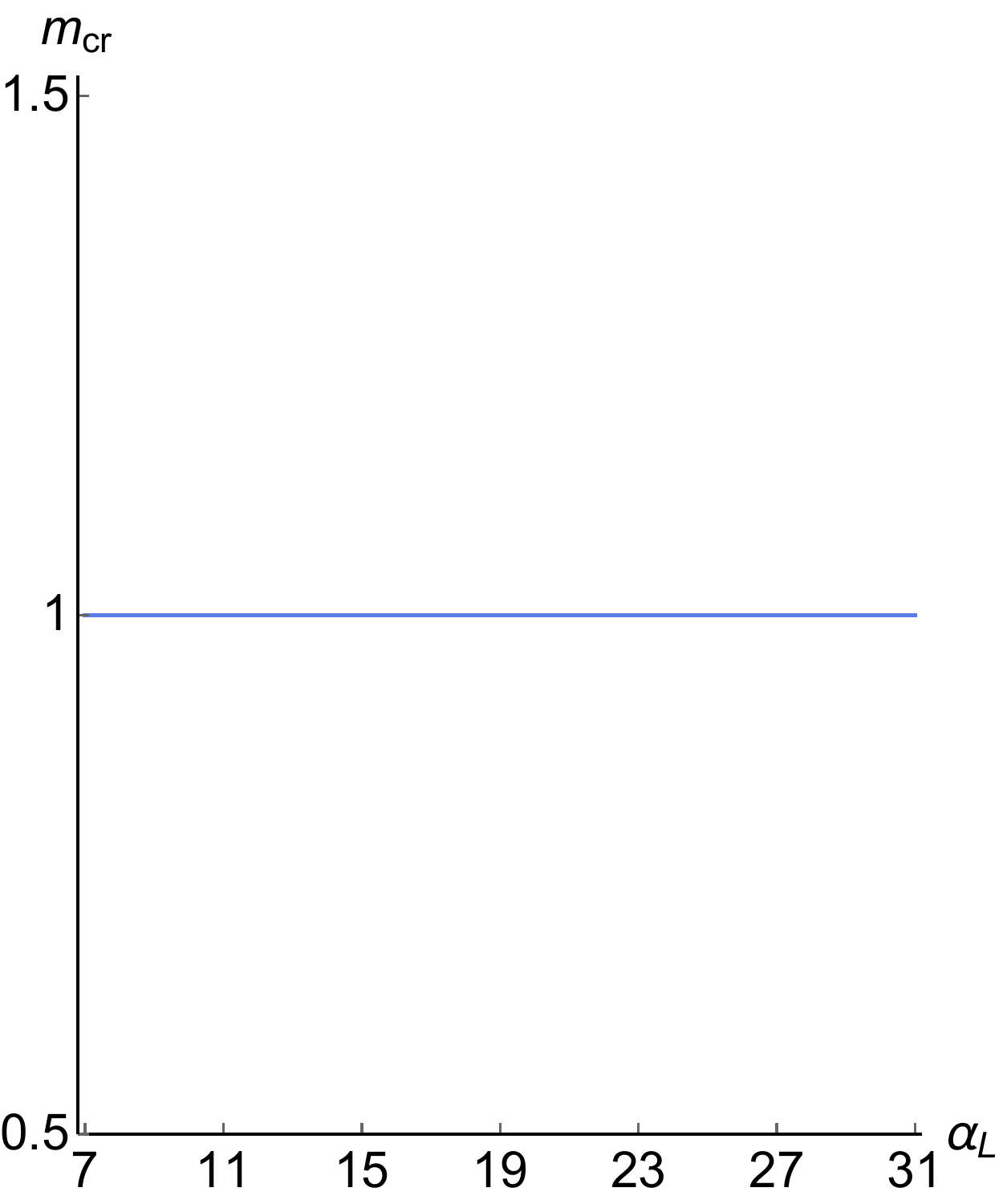}
		\caption{$\alpha_R = 2.85$}
		\label{fig:mvsalphaL_mcr1}
	\end{subfigure}
	\begin{subfigure}{.32\linewidth}
		\centering
		\includegraphics[width=\textwidth]{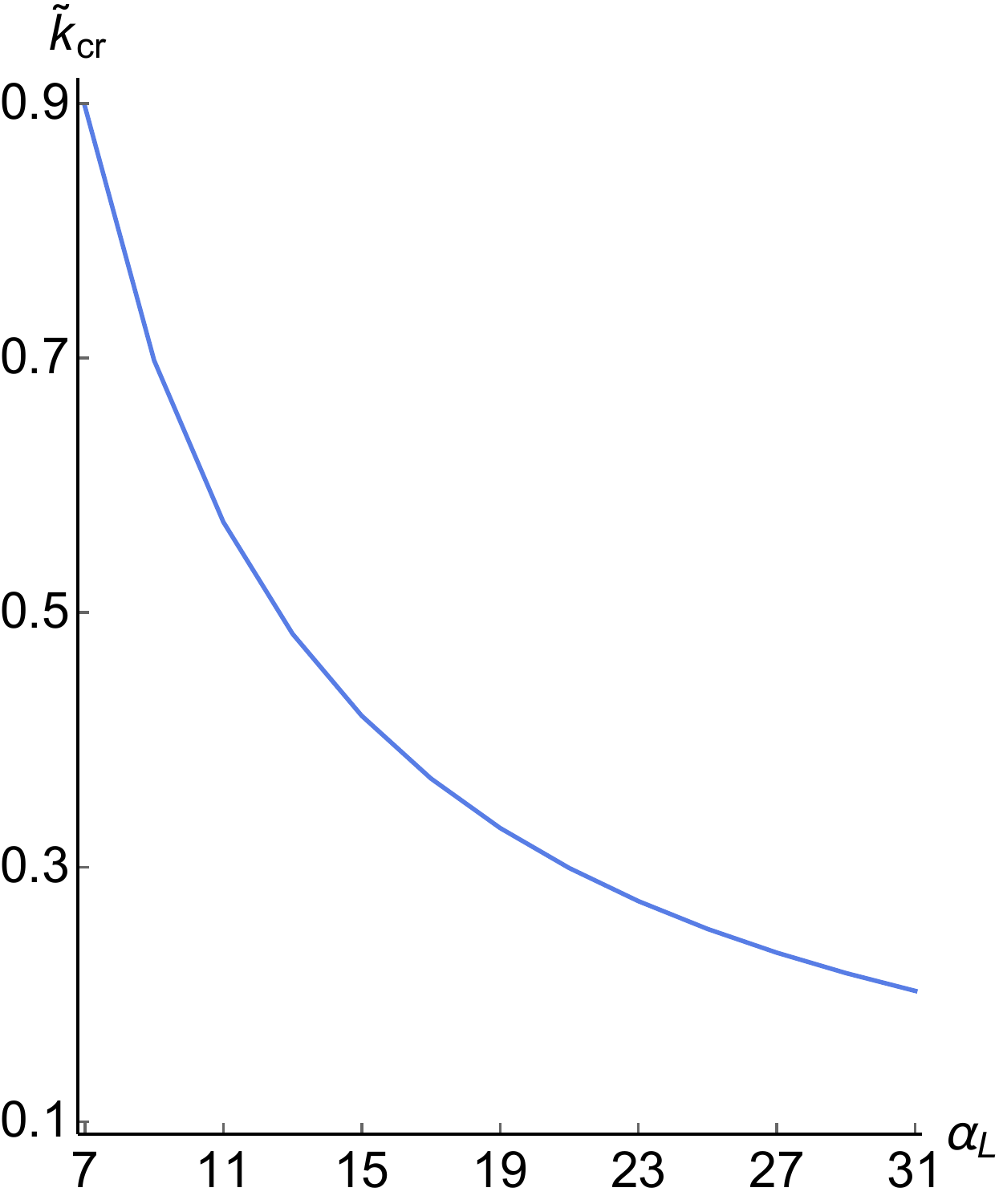}
		\caption{$\alpha_R = 2.85$}
		\label{fig:kzvsalphaL_mcr1}
	\end{subfigure}
	\begin{subfigure}{.32\linewidth}
		\centering
		\includegraphics[width=\textwidth]{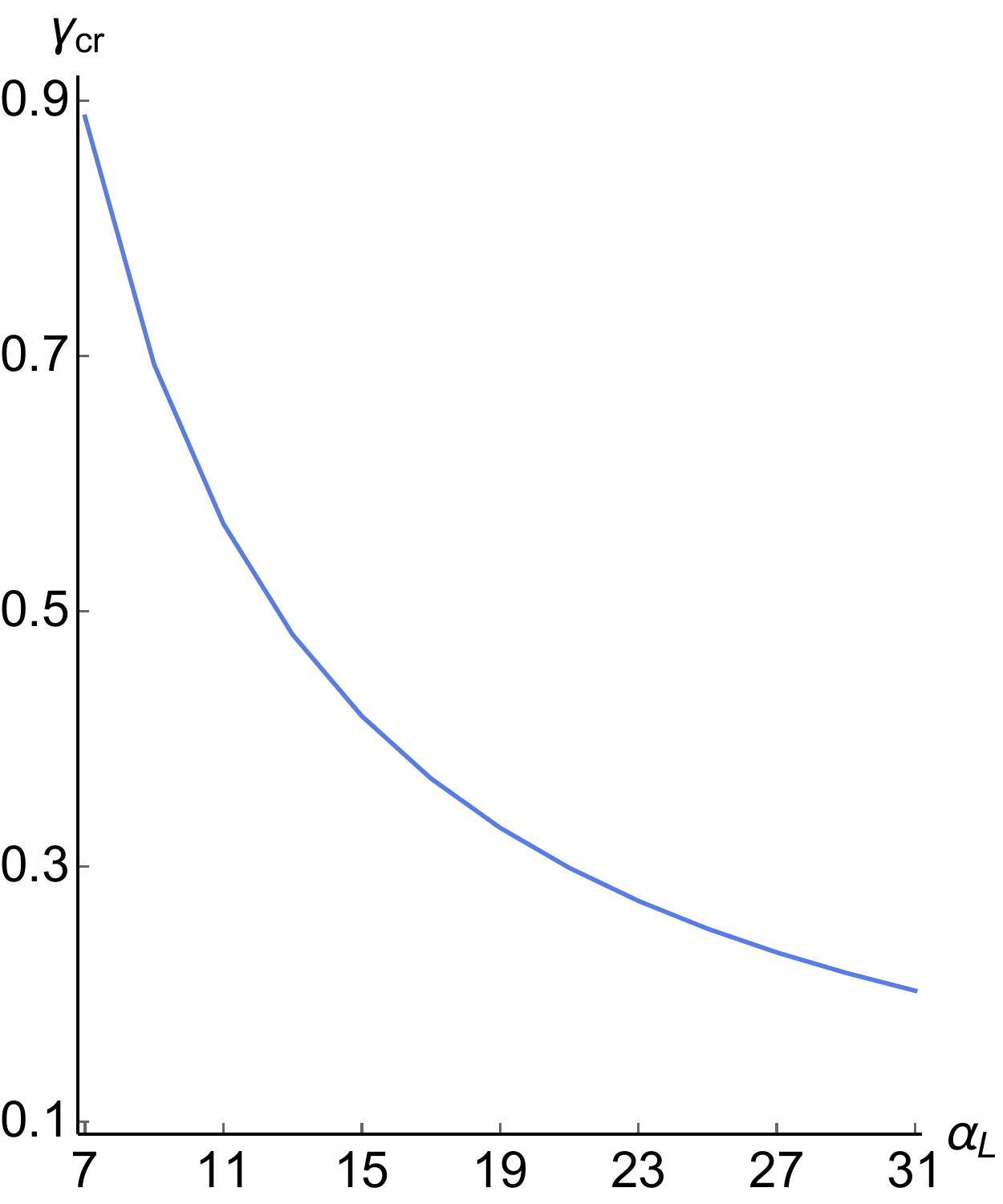}
		\caption{$\alpha_R = 2.85$}
		\label{fig:gammavsalphaL_mcr1}
	\end{subfigure}
	\caption{Plot of the critical values versus $\alpha_L$, fixing the initial aspect $\alpha_R= 2.85$. (a) The critical circumferential wavenumber is $m_{\rm cr} = 1$, since $\alpha_L$ is big enough. (b) Plot of the axial critical wavenumber $\tilde{k}_{\rm cr}$ and (c) of the marginal stability threshold $\gamma_{\rm cr}$ versus $\alpha_L$}
	\label{fig:vsalphaL_mcr1}
\end{figure}

The parameter $\alpha_L$ also affects the marginal stability thresholds of  the HT.  In Fig. \ref{fig:vsalphaL_mcr1}, we fix $\alpha_R = 2.85$, i.e. $m_{\rm cr} = 1$ as we can see in Fig. \ref{fig:mvsalphaL_mcr1}, and we plot the marginal stability threshold $\gamma_{\rm cr}$ and the axial critical wavenumber $\tilde{k}_{\rm cr}$ versus $\alpha_L$. Both the two curves, Figs. \ref{fig:kzvsalphaL_mcr1} - \ref{fig:gammavsalphaL_mcr1} are coherent with the typical mechanical behaviour of a twisted Euler rod: as we increase the length as the cylinder instabilizes and the emerging pattern is an helix of pitch $1/\tilde{k}$ \cite{green1958stability,gent2004torsional}.

Figs. \ref{fig:vsalphaR}-\ref{fig:vsalphaL_mcr1} characterize the loss of marginal stability as a function of the thickness ratio $\alpha_R$ and the slenderness ratio  $\alpha_L$. By using these characteristic values, in the next section we study the development of the looped configuration far beyond the marginal stability threshold: we implement a finite element code to discretize and numerically solve the fully nonlinear BVP given by Eqs. \eqref{eq:balance}-\eqref{eq:noloadsP}.

\section{Numerical simulations}
\label{sec:numerical_simulation_clooping}
\begin{figure}[h!]
	\begin{subfigure}{.43\linewidth}
		\centering
		\includegraphics[width=\textwidth]{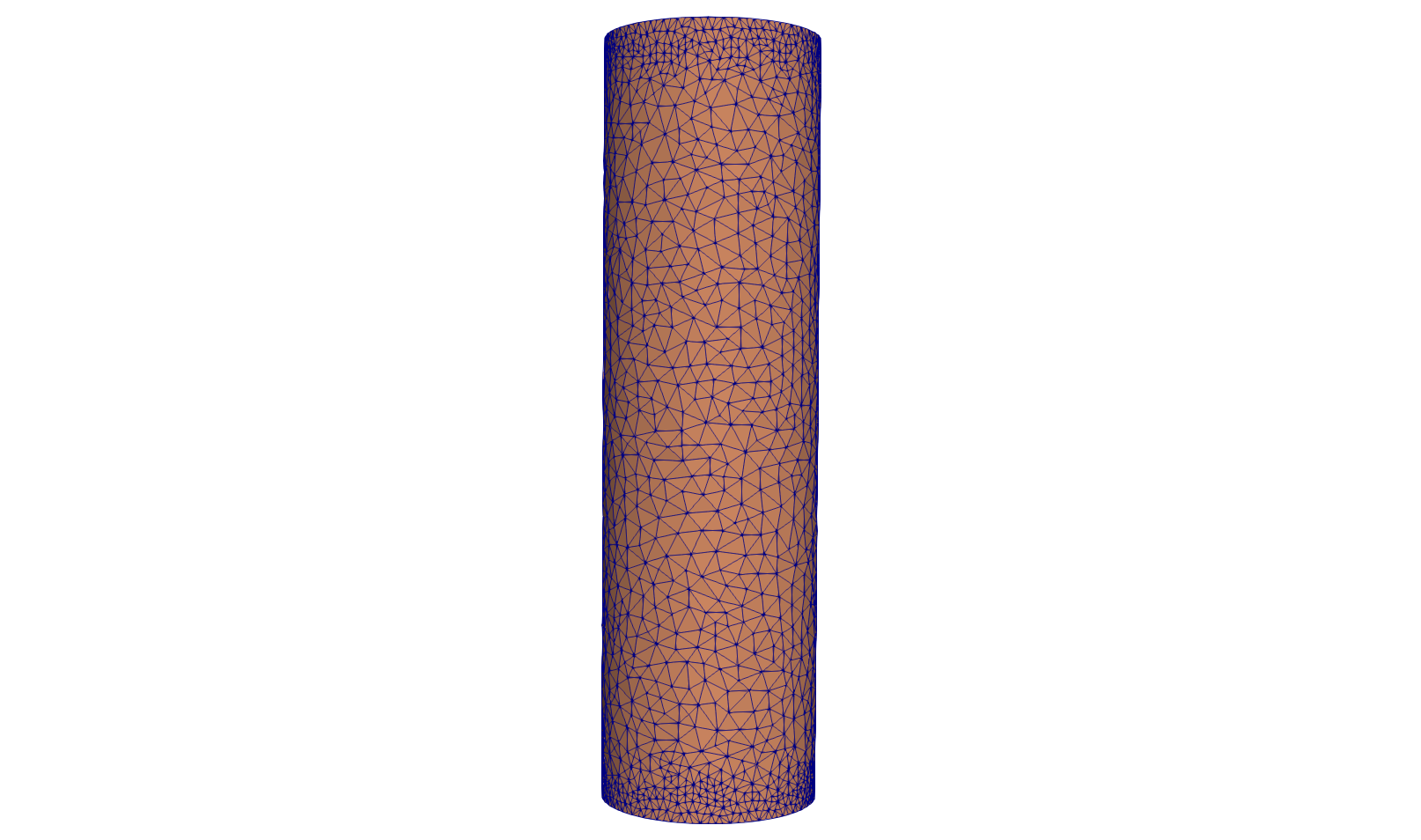}
		\caption{Mesh}
		\label{fig:mesh_front}
	\end{subfigure}
	\begin{subfigure}{.43\linewidth}
		\centering
		\includegraphics[width=\textwidth]{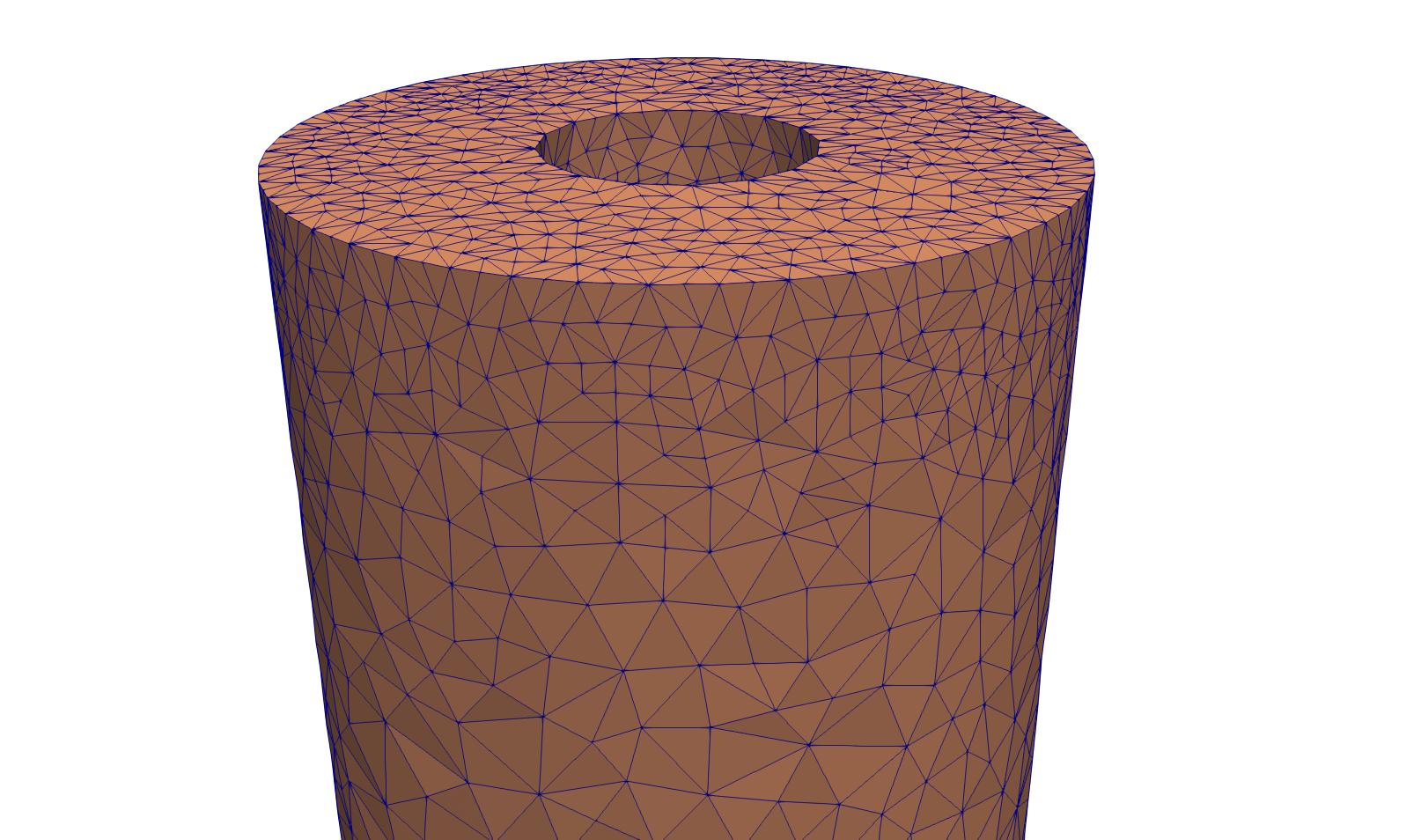}
		\caption{Top view}
		\label{fig:mesh_top}
	\end{subfigure}%
	\caption{Mesh generated through MSHR for $\alpha_R = 2.85$ with local refinement.
	}
	\label{fig:mesh}
\end{figure}

The boundary value problem is implemented by using the open source tool for solving partial differential equations FEniCS \cite{logg2012automated}. We generate as a computational domain a hollow cylinder with a non-structured tetrahedral mesh created through the module MSHR \cite{logg2012automated}, see Fig. \ref{fig:mesh}, where we refine the mesh around the two bases $z = 0$ and $z = \alpha_L$, see Fig. \ref{fig:mesh_top}.
Since the cylinder is finite, to solve the non linear elastic BVP Eqs.~\eqref{eq:incompressibility}, \eqref{eq:balance} and \eqref{eq:noloadsP}, we need to impose suitable boundary conditions on the top and bottom bases. First, we assume a stronger form of  \eqref{eq:zL}, imposing
\begin{equation}
\label{eq:bcbasi_u0}
\vect{u} = 0 \quad \hbox{on } z = 0,\,  z = \alpha_L.
\end{equation}
We discretize the computational domain by using a mesh composed by $78338$ elements and to solve the problem, we use the Taylor-Hood $\vect{P_2}-P_1$ element, i.e. the displacement field is given by a continuous, piecewise quadratic function while the pressure field by a continuous, piecewise linear function. The choice of this particular element is motivated by its stability for non-linear elastic problems \cite{auricchio2013approximation}. We introduce an additional term to stabilize the numerical scheme which penalizes the volumetric deformations, since at the discrete level and the use $\vect{P_2}$-elements can result in unphysical values for the determinant of $\tens{F}_{\rm e}$ \cite{auricchio2013approximation}. Second, in order to evaluate the effect of the boundary conditions on the top and the bottom surfaces, we perform further simulations imposing a weaker form of  \eqref{eq:zL}, precisely
\begin{equation}
\label{eq:bcuz0}
\left\{
\begin{aligned}
&	\vect{u} \cdot \vect{e}_{r} = 0&&\hbox{on } z = 0,\,  z = \alpha_L,\\
&\vect{u} \cdot \vect{e}_{\vartheta} = 0&&\hbox{on } z = 0,\,  z = \alpha_L,\\
&\int_{z = i} u_{z} = 0 &&\hbox{where } i = 0, \,\alpha_L.
\end{aligned}
\right.
\end{equation}
In this case, the mesh is composed by $78426$ elements. We solve the discretised form of the equilibrium equation Eq. \eqref{eq:balance} in the Lagrangian form using
a Newton method. The control parameter $\gamma$ is incremented of $\delta \gamma$ when the Newton method converges, the numerical solution is used as initial guess for the following Newton cycle. The increment $\delta \gamma$ is automatically reduced near the theoretical marginal stability threshold and when the Newton method does not converge. The numerical simulation is stopped when $\delta \gamma< 10^{-6}$. To trigger the mechanical instability, a small perturbation of an amplitude of $10^{-5}$, having the shape of the critical mode computed in Section \ref{sec:lin_stab_clooping}, is applied at the free boundary of the mesh.
The numerical algorithm is implemented in Python through the open-source computing platform FEniCS (version 2018.1) \cite{logg2012automated}.  We use PETSc \cite{balay2018petsc} as linear algebra back-end and MUMPS \cite{Amestoy_2000} as linear solver.

\subsection{Simulation results}
\label{subsec:result_sim}
In this section, we discuss the results of the numerical simulations varying the two physical parameters $\alpha_R$ and $\alpha_L$  firstly within the biological range given by the experimental papers, and secondly outside this range to further understand the role of geometrical parameters in pattern selection, validating the results against some known features in the limiting case of solid cylinder.
\subsection{Biological range of dimensionless parameters}
\begin{figure}[h!]
	\centering
	\includegraphics[width=\textwidth]{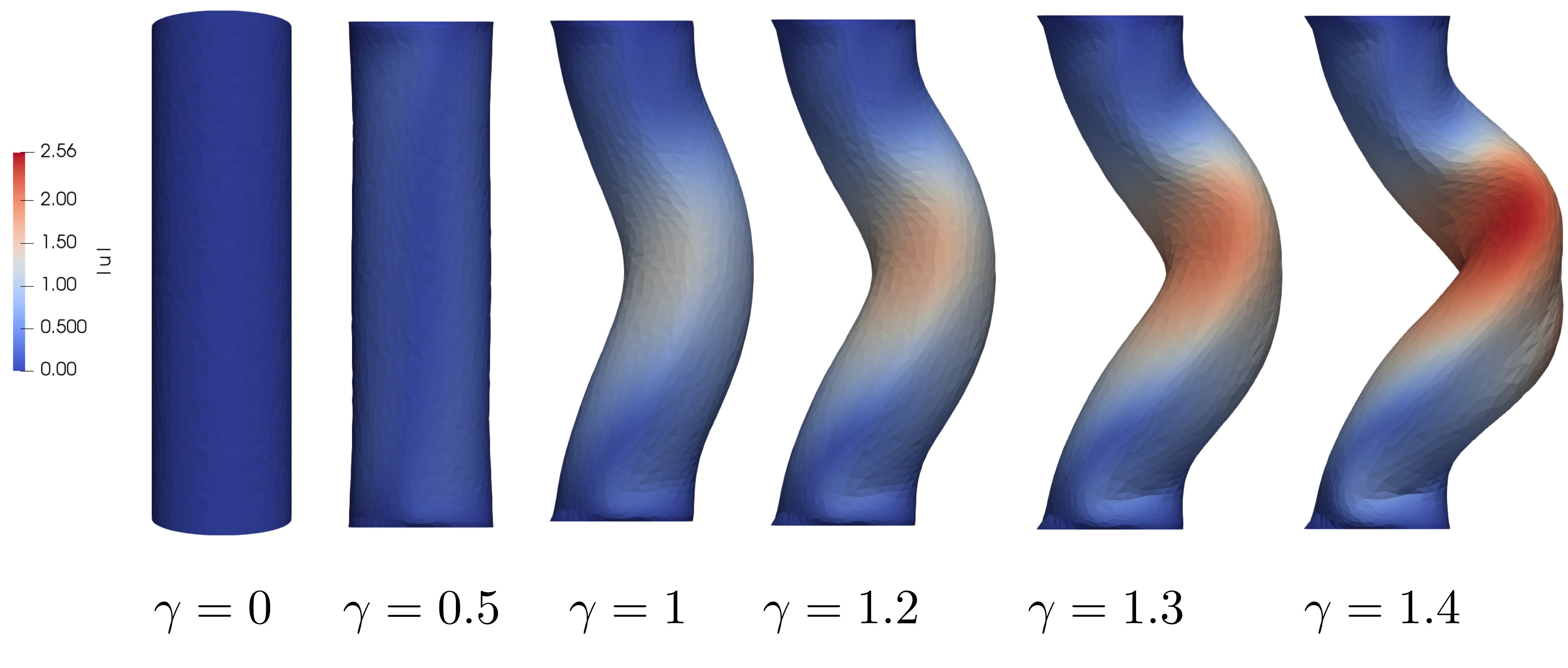}
	\caption{Actual configuration of the buckled tube when $\alpha_R = 2.85$ and $\alpha_L = 7$. In such conditions $\gamma_{\rm cr} \simeq 0.881748$.}
	\label{fig:sim285}
\end{figure}

In Fig. \ref{fig:sim285}, we plot the looping development of the HT for a hollow cylinder with $\alpha_R = 2.85$, $\alpha_L = 7$,  imposing Eq. \eqref{eq:bcbasi_u0} on the top and bottom bases. We show the actual configuration for several values of the control parameter $\gamma$. As $\gamma$ increases, the cylinder displays an helical pattern and the simulation stops around $\gamma \simeq 1.4$ because the inner lumen closes and its internal surface self-contact.

\begin{figure}[h!]
	\begin{subfigure}{.59\linewidth}
		\centering
		\includegraphics[width=0.83\textwidth]{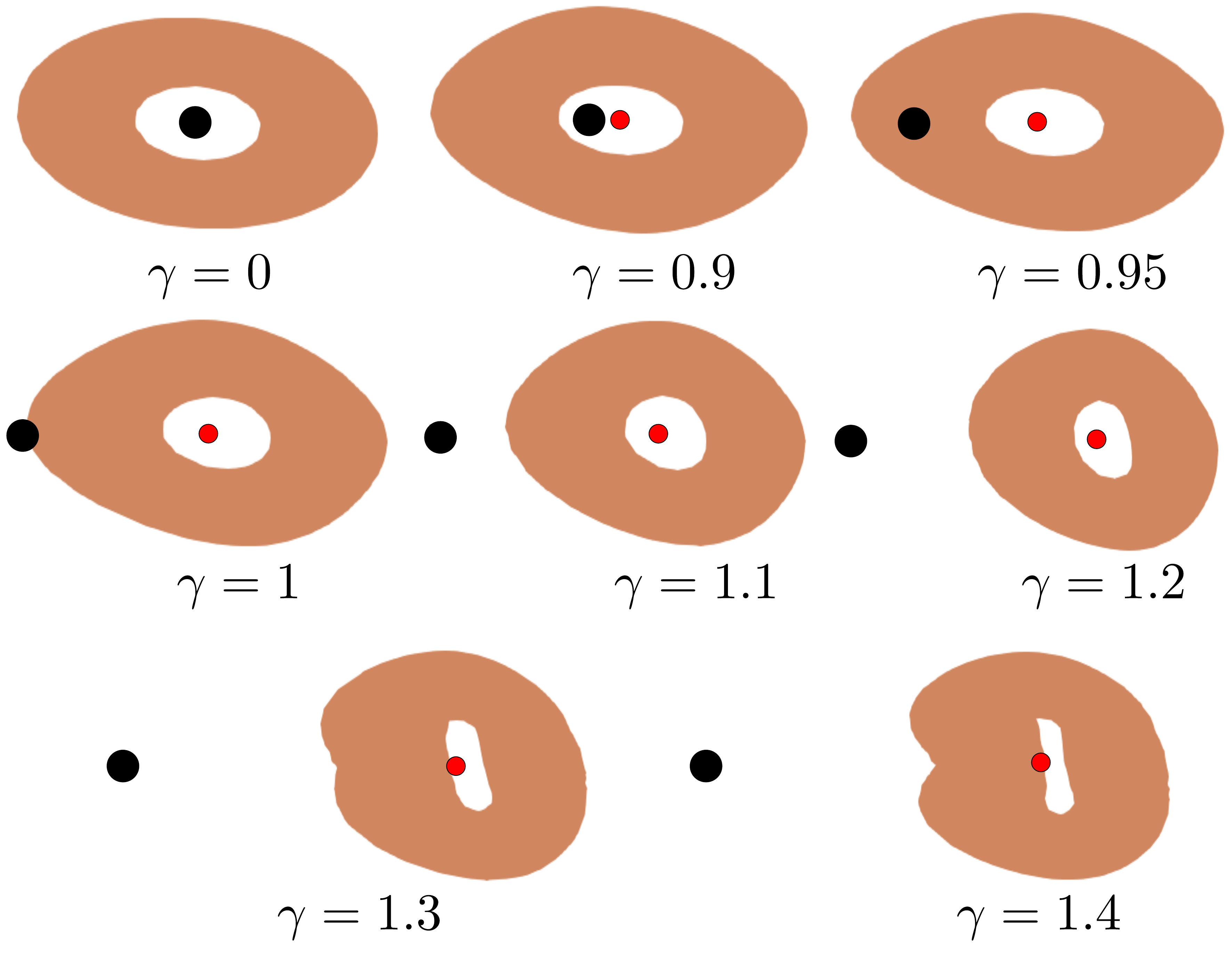}
		\caption{$z = 3.5$}
		\label{fig:shape_nonlineare}
	\end{subfigure}
	\begin{subfigure}{.39\linewidth}
		\centering
		\includegraphics[width=0.85\textwidth]{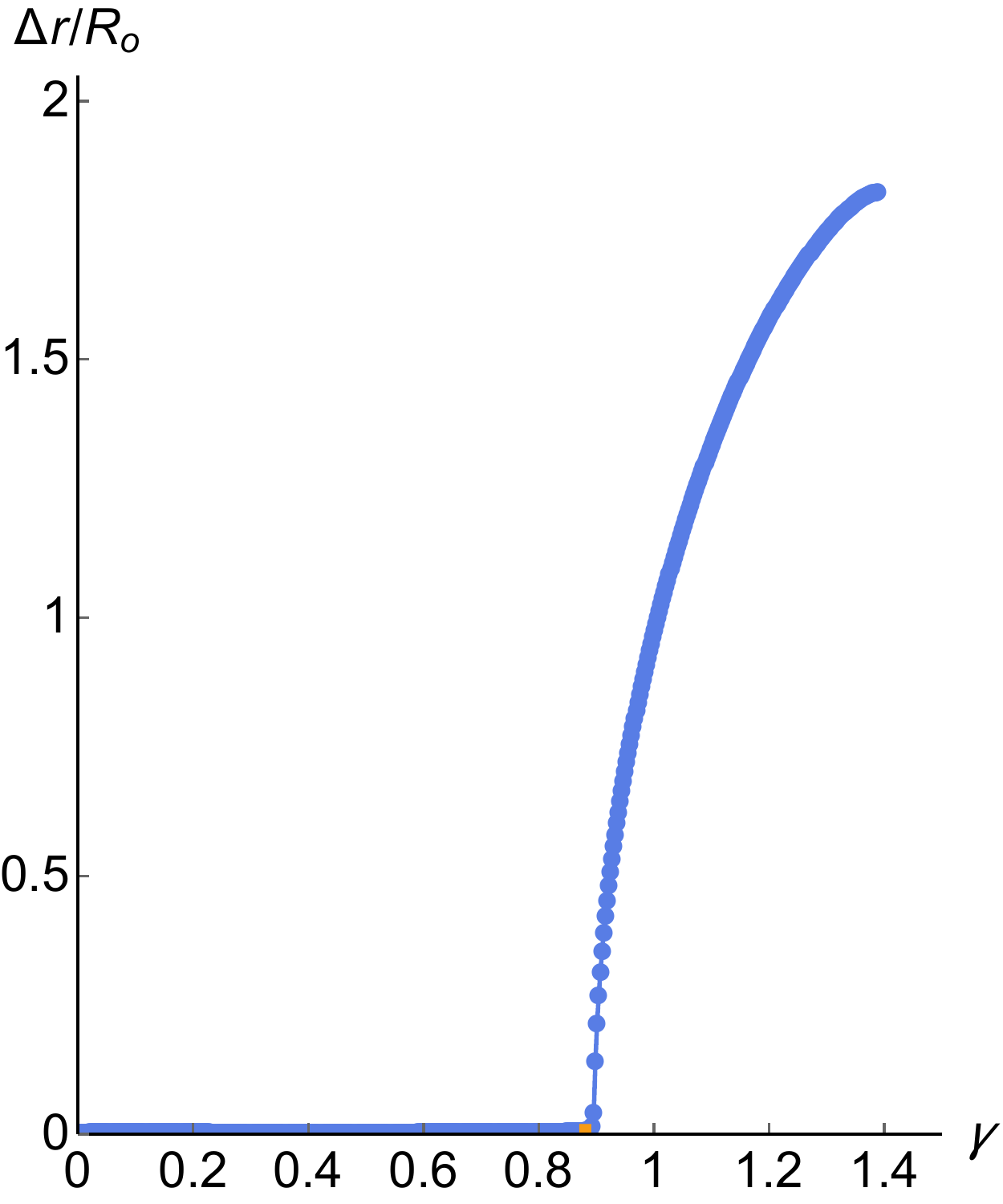}
		\caption{$z = 3.5$}
		\label{fig:z35}
	\end{subfigure}
	\caption{(a) Plot of the different shapes of the mid-section $z = \alpha_L/2 = 3.5$ at different values of the control parameter $\gamma$. The black dot is the point on the $z$-axis $O = (0,0,\alpha_L/2)$, while the red one is the centroid of the section which moves away form $O$. The shape of the section changes and the inside hole is getting smaller causing the stop of the numerical simulation.
		(b) Bifurcation diagram where we show the dimensionless parameter $\Delta r/ R_{\rm o}$ defined in Eq. \ref{eq:delta_r} versus the control parameter $\gamma$ when $\alpha_R = 2.85$ and $\alpha_L = 7$. The numerical simulation is validated against the marginal stability threshold computed with the linear stability analysis (orange square, $\gamma_{\rm cr} \simeq 0.881748$).}
	\label{fig:nonlineare}
\end{figure}

The  closure of the internal hole can be visualized from Fig. \ref{fig:nonlineare}, where we  depict the mid-sectional cut at $z = \alpha_L/2 = 3.5$. At the beginning $\gamma = 0$, the central section looks like a circular crown, while as the control parameter $\gamma$ increases, as the internal hole reduces, the circle stretches to become an ellipse, see the last frame of Fig. \ref{fig:shape_nonlineare} when $\gamma = 1.4$.

In order to study the amplitude of the helical pattern, we cut the cylinder with a plane $z = c$, where $c$ is a constant and it is perpendicular to the $z$-axis. Fixing $\alpha_R = 2.85$ and $\alpha_L = 7$, we choose a particular and significant value of $c$ respecting the symmetry of the system, i.e. we select the central section $c = \alpha_L /2 = 3.5$. Hence, we define  
\begin{align}
\label{eq:delta_r}
\Delta r = {\rm dist} (C_c-O),
\end{align}
where $C_c$ is the centroid of the considered section, the red dot in Fig. \ref{fig:shape_nonlineare} and $O = (0,0,c)$, the black dot in Fig. \ref{fig:shape_nonlineare}.   In Fig. \ref{fig:z35}, we plot $\Delta r / R_{\rm o}$ versus $\gamma$ to measure the distance of the centroid of the section from the $z$-axis. We observe that there is an excellent agreement with the marginal stability thresholds computed in the previous section, verifying  the results obtained by the numerical code against the theoretical predictions. Both bifurcation diagrams exhibit a continuous transition from the unbuckled to the buckled configuration, displaying the typical behaviour of a supercritical pitchfork bifurcation.

\begin{figure}[h!]
	\centering
	\includegraphics[width=0.7\textwidth]{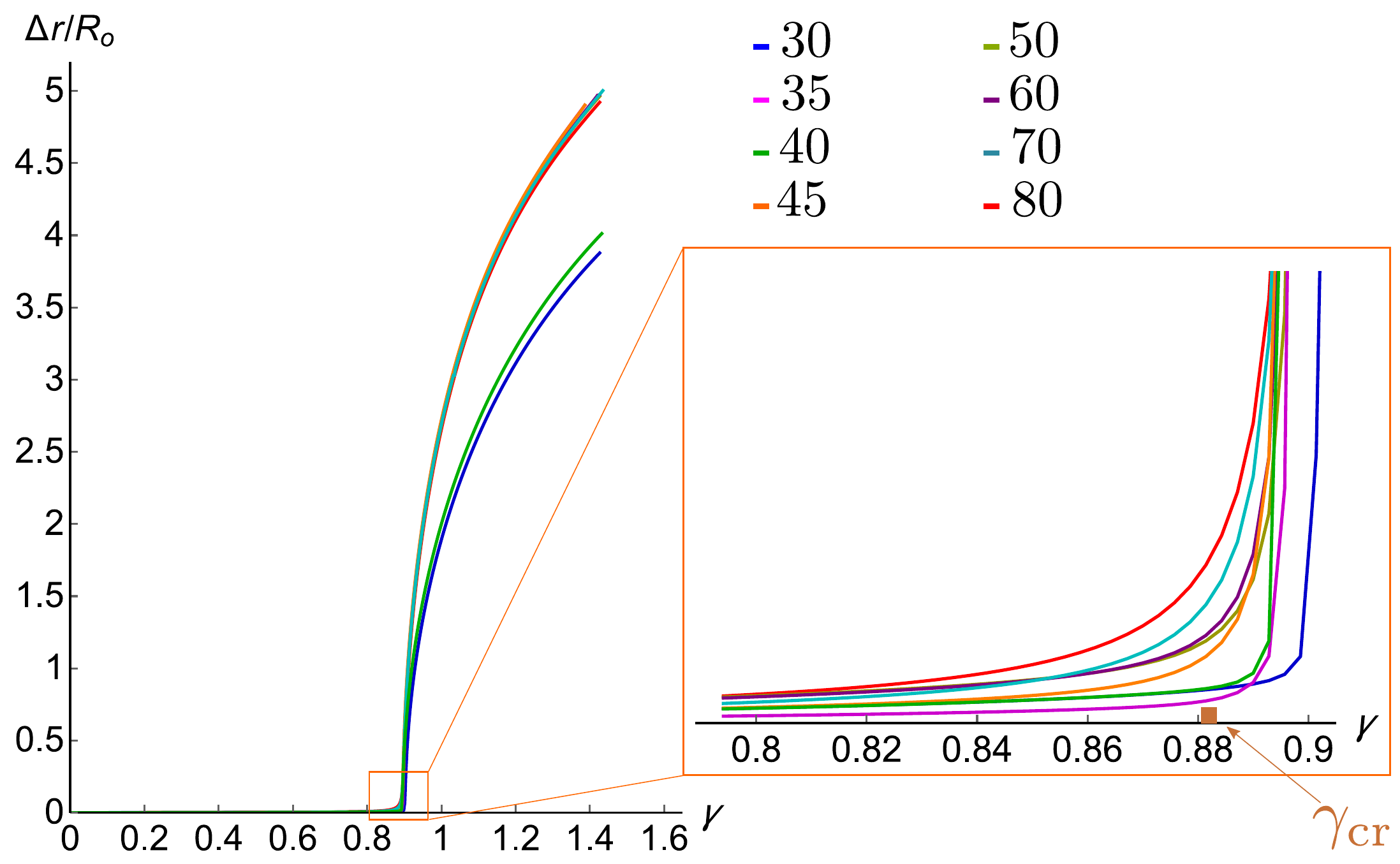}
	\caption{Plot of the bifurcation diagram varying the numbers of faces $30,35,40,45,50,60,70,80$ on the side of the cylinder \cite{logg2012automated}. In the inset, we highlight the behaviour near the marginal stability threshold.}
	\label{fig:conv_analisis}
\end{figure}

\begin{figure}[h!]
	\centering
	\includegraphics[width=0.85\textwidth]{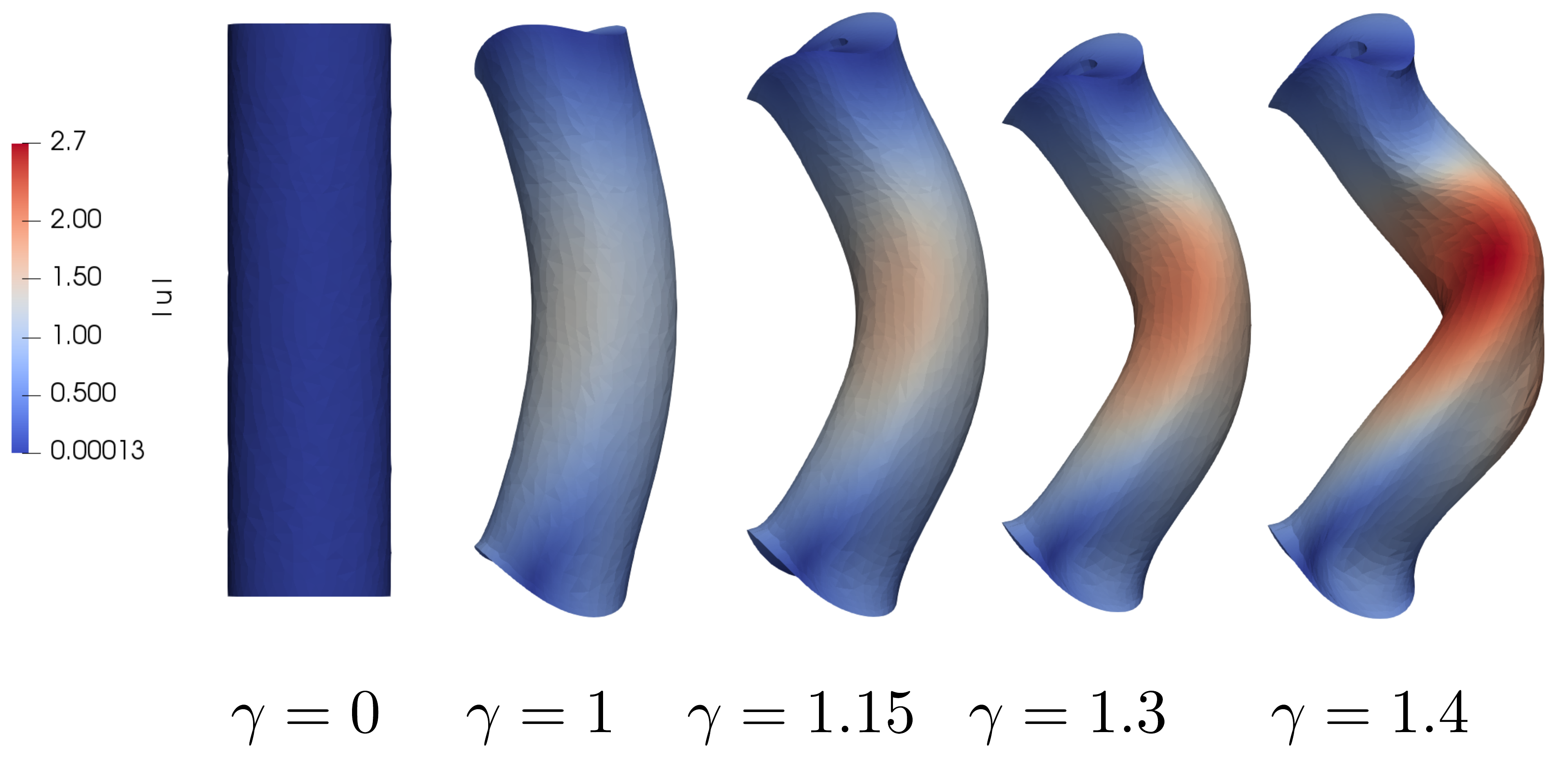}
	\caption{Actual configuration of the looped tube when $\alpha_R = 2.85$ and $\alpha_L = 7$. In such conditions $\gamma_{\rm cr} \simeq 0.881748$.}
	\label{fig:sim285_changeBC}
\end{figure}

\begin{figure}[h!]
	\begin{subfigure}{.49\linewidth}
		\centering
		\includegraphics[width=0.9\textwidth]{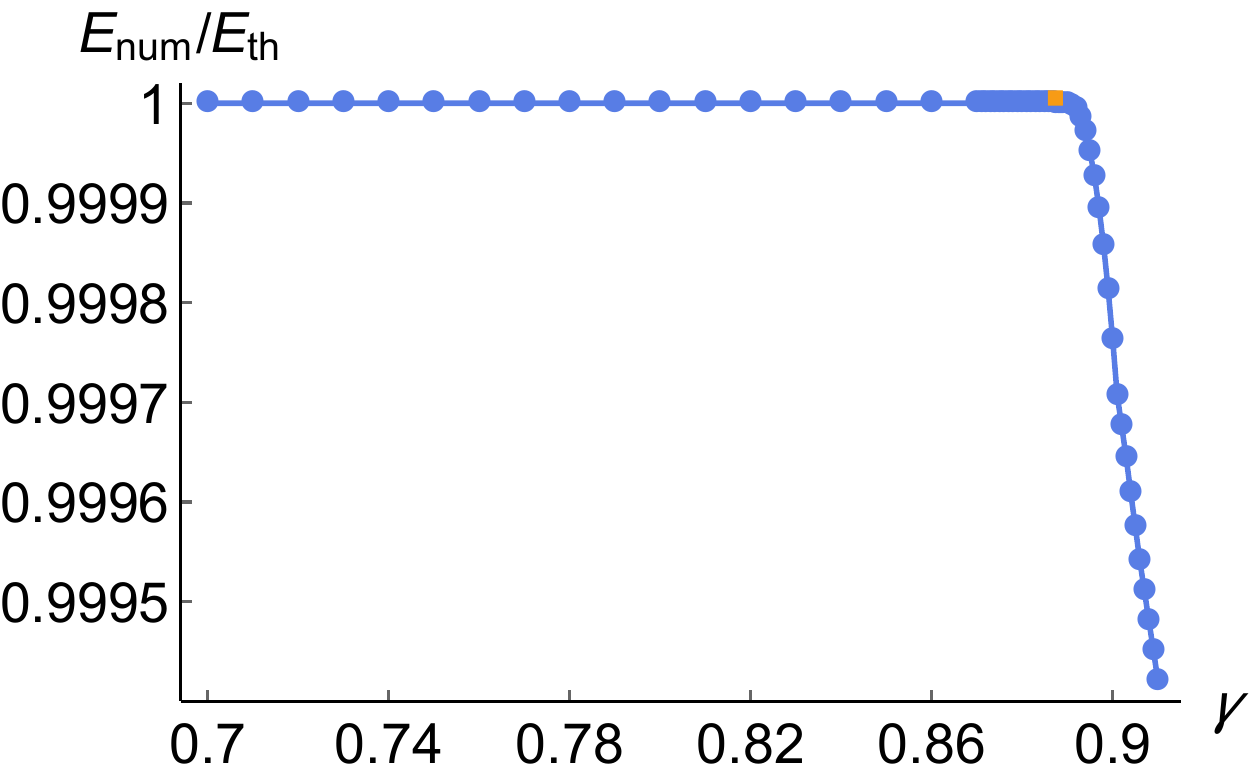}
		\caption{$\vect{u}=0$ as BC}
		\label{fig:ener_u0}
	\end{subfigure}
	\begin{subfigure}{.49\linewidth}
		\centering
		\includegraphics[width=0.9\textwidth]{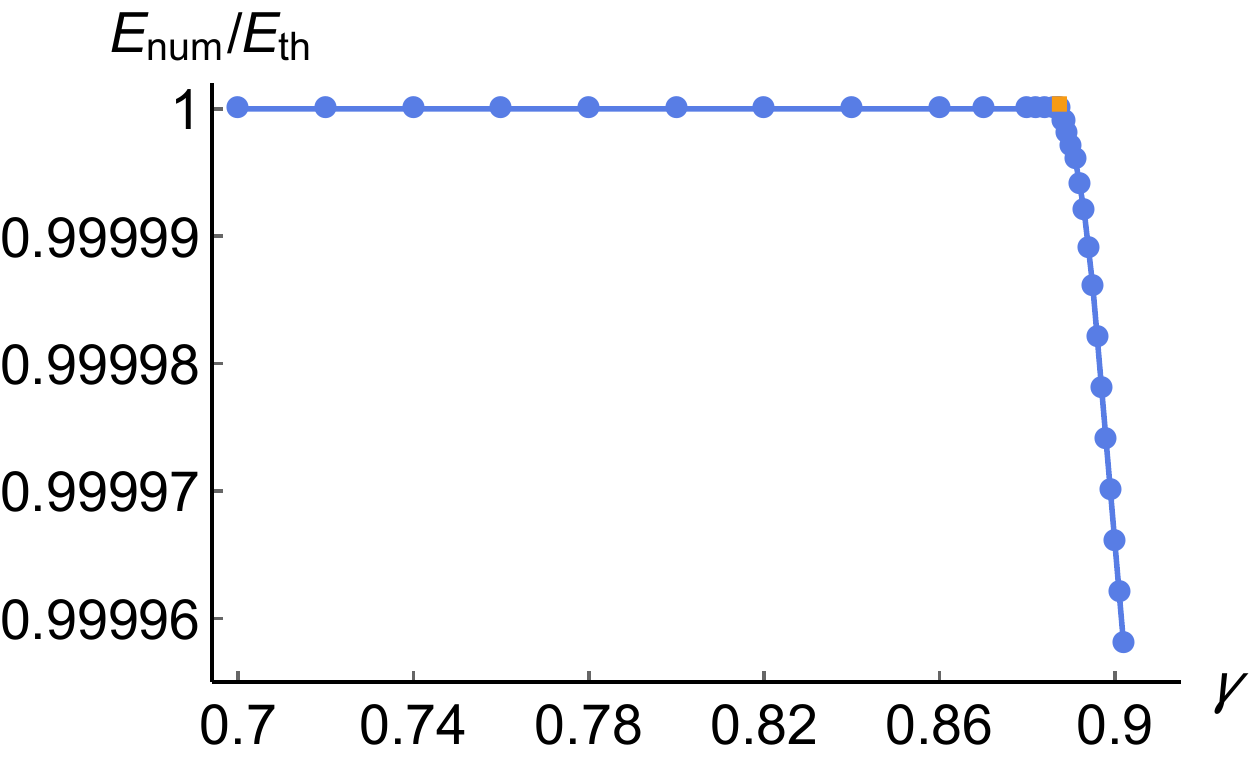}
		\caption{$\int_{S_{i}}u_z = 0$ as BC}
		\label{fig:ener_changeBC}
	\end{subfigure}%
	\caption{Plot of the ratio $E_{\rm num}/E_{\rm th}$ for (a) $\alpha_R = 2.85$ and $\vect{u} = 0$ as boundary condition and (b) $\alpha_R = 2.85$ and $\int_{S_{i}}u_z = 0$ with $i$ equal to the top and the bottom bases. The orange square denotes the theoretical marginal stability threshold computed as  in Section \ref{sec:lin_stab_clooping}.}
	\label{fig:ener}
\end{figure}

Finally, in Fig. \ref{fig:conv_analisis}, we perform the convergence analysis on our numerical simulations. We run several simulations at fixed geometry whilst increasing the number of the tetrahedra in the mesh, precisely we vary the numbers of faces on the side of the cylinder \cite{logg2012automated}. As the mesh gets finer and finer, as all the curves collapse on the same  and the numerical instability threshold is delayed for a coarser mesh, as expected due to the lower accuracy of the numerical approximation.
In Fig. \ref{fig:sim285_changeBC}, we fix the same geometry as in Fig. \ref{fig:sim285}, i.e. $\alpha_R = 2.85$ and $\alpha_L = 7$, we solve again the fully nonlinear BVP Eqs.~\eqref{eq:incompressibility}, \eqref{eq:balance} and \eqref{eq:noloadsP}, but we change the boundary condition on the two bases, substituting Eq. \eqref{eq:bcbasi_u0} with Eq. \eqref{eq:bcuz0}. As in the previous case, the simulation stops because the hole closes and the internal surfaces enters in self-contact  approximately  at the same value of the control parameter $\gamma \simeq 1.4$ as in the previous case. % The scale-bar of the displacement does not start with zero as the lower value since on the two bases the displacement is not exactly zero.
Thus, the two different choices of the boundary conditions for the top and bottom ends of the HT result in a small deviation of the onset of the bifurcation. In Fig. \ref{fig:ener}, we plot the ratio $E_{\rm num}/E_{\rm th}$, where $E_{\rm num}$ is the energy computed through the finite elements computation while $E_{\rm th}$ is the one of the axis-symmetric solution, versus the control parameter $\gamma$ fixing $\alpha_R = 2.85$ and $\alpha_L = 7$ for the two different boundary conditions on the two bases, respectively Eq. \eqref{eq:bcbasi_u0} - \eqref{eq:bcuz0}. As expected, the buckled configuration exhibits  in both cases a total mechanical energy lower than the one in the unbuckled state. In  Fig. \ref{fig:ener_u0}, we fixed the whole displacement field $\vect{u}= 0$, and the instability is a bit delayed with respect to the theoretical marginal stability threshold $\gamma_{\rm cr}$. In Fig. \ref{fig:ener_changeBC} we require that the displacement along $z$ is zero in a weaker way, and the threshold is closer to the theoretical prediction. Furthermore we observe that the energy decays continuously as $\gamma$ grows beyond thresholds, confirming that the bifurcation is in both cases supercritical, see Figs. \ref{fig:nonlineare} and \ref{fig:conv_analisis}.

\subsection{Geometrical effects and solid cylinder limit}
We now investigate how the geometry, i.e. varying $\alpha_R$ and $\alpha_L$, influences the buckled configuration in a wider range of dimensionaless parameters, extending the linear stability analysis for some cases illustrated in Section \ref{sec:lin_stab_clooping}.

\begin{figure}[h!]
	\centering
	\includegraphics[width=0.75\textwidth]{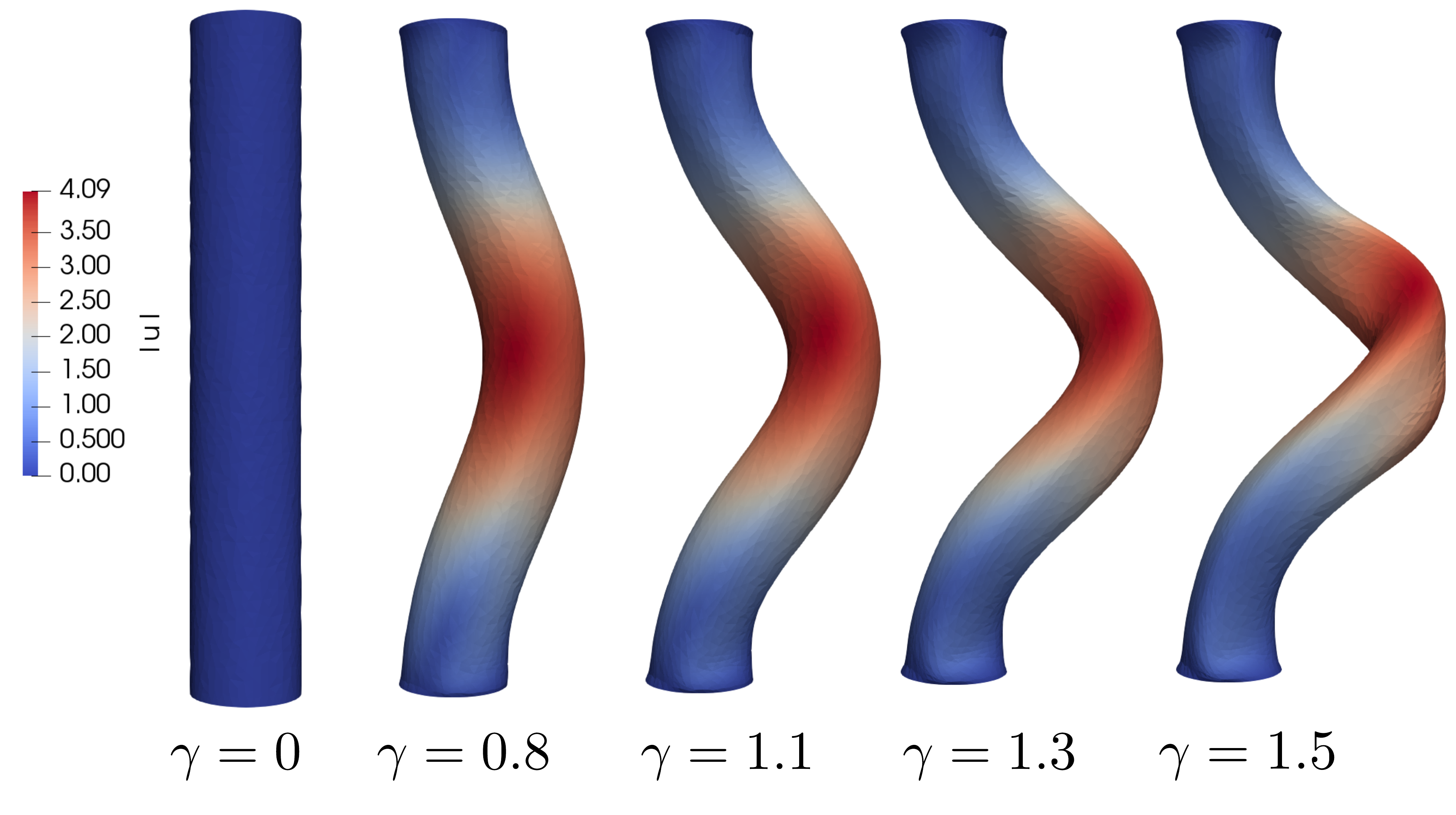}
	\caption{Actual configuration of the buckled tube for when $\alpha_R = 2.85$ and $\alpha_L = 12$. In such conditions $m_{\rm cr} = 1$, $\tilde{k}_{\rm cr} = 2 \pi / \alpha_L$ and $\gamma_{\rm cr}=0.521594$.}
	\label{fig:alphaL12}
\end{figure}

First, keeping fixed $\alpha_R = 2.85$, we increase the other dimensionless quantity, i.e. setting $\alpha_L = 12$. In Fig. \ref{fig:alphaL12}, we plot the buckled configuration of the cylinder having imposed Eq. \eqref{eq:bcbasi_u0} on the two bases. We show the actual configuration for several values of the control parameter $\gamma$. As $\gamma$ increases, as the cylinder displays an helical pattern with a higher amplitude with respect to the one with $\alpha_L = 7$, see the scale-bar of Fig. \ref{fig:alphaL12}.

\begin{figure}[h!]
	\begin{subfigure}{.47\linewidth}
		\centering
		\includegraphics[width=0.9\textwidth]{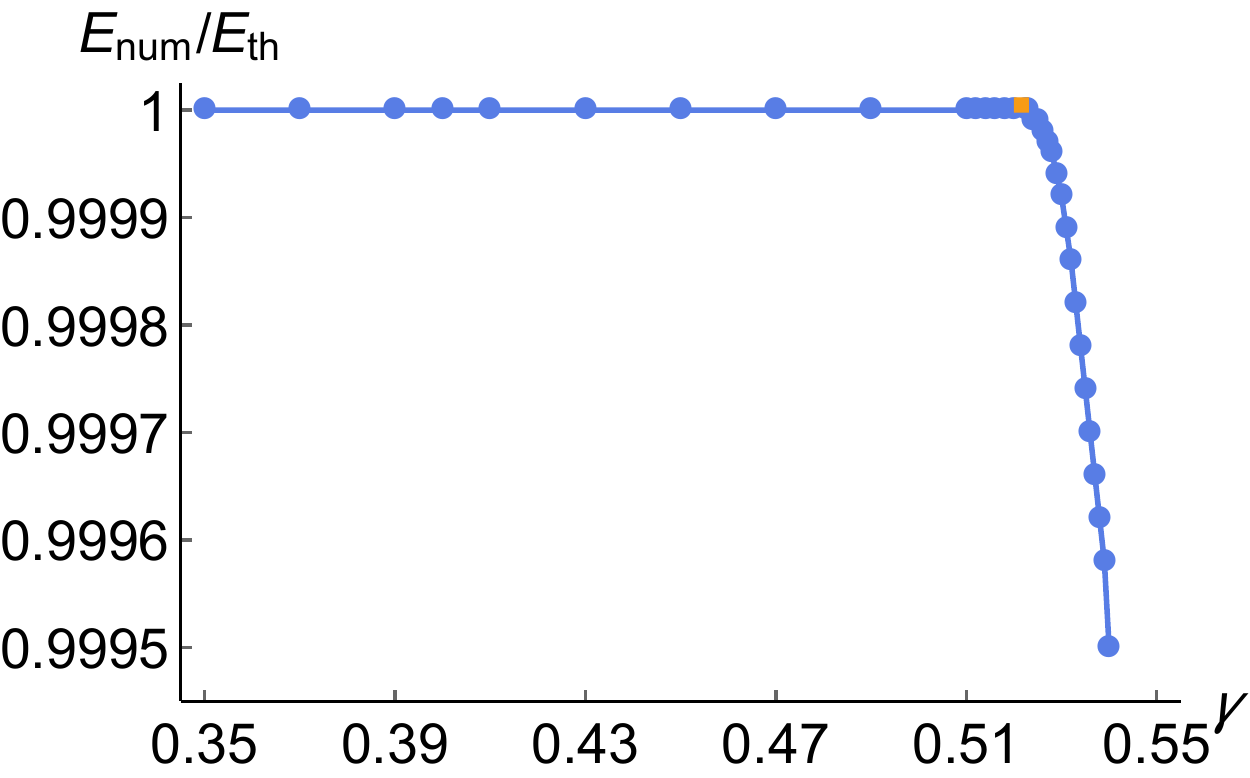}
		\caption{$\alpha_L = 12$, $\vect{u}=0$ as BC}
		\label{fig:ener_u0_L12}
	\end{subfigure}
	\begin{subfigure}{.47\linewidth}
		\centering
		\includegraphics[width=0.65\textwidth]{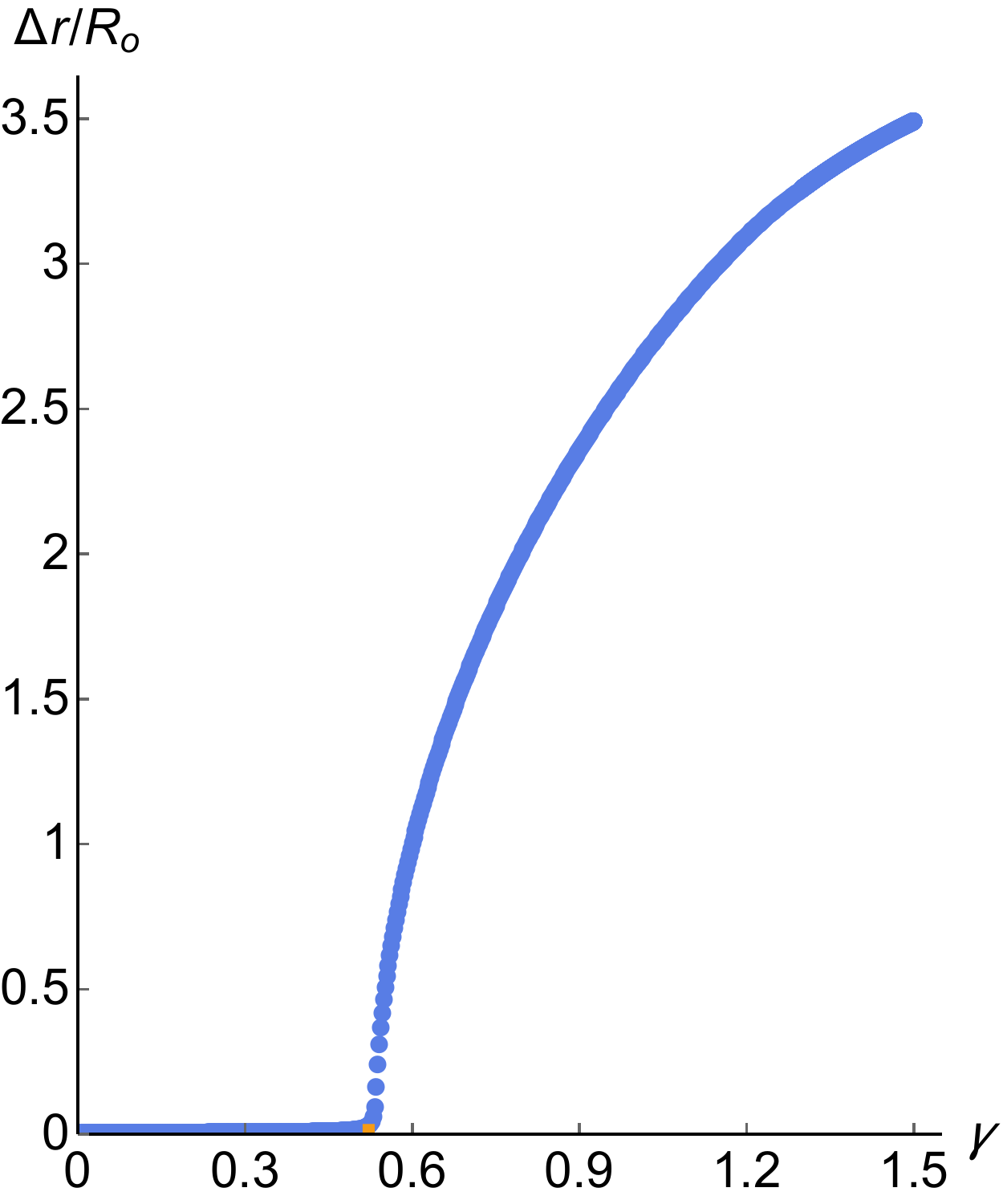}
		\caption{$\alpha_L = 12$, $z = \alpha_L/2$}
		\label{fig:biforcazione_L12}
	\end{subfigure}%
	\caption{(a) Plot of the ratio $E_{\rm num}/E_{\rm th}$ for $\alpha_R = 2.85$, $\alpha_L = 12$ and $\vect{u} = 0$ as boundary condition. (b) Bifurcation diagram where we show the dimensionless parameter $\Delta r/ R_{\rm o}$ versus the control parameter $\gamma$ when $\alpha_R = 2.85$, $\alpha_L = 12$ and $z = \alpha_L/2 = 6$. The orange square denotes the theoretical marginal stability threshold computed as exposed in Section \ref{sec:lin_stab_clooping}, i.e. $m_{\rm cr} = 1$, $\tilde{k}_{\rm cr} = 2\pi/\alpha_L$ and $\gamma_{\rm cr} = 0.521594$.
	}
	\label{fig:L12}
\end{figure}

Using a mid-section  perpendicular to its axis, we plot in Fig. \ref{fig:biforcazione_L12} the distance of the centroid of this section from $O = (0,0,6)$, i.e. $\Delta r/R_{\rm 0}$ versus the control parameter $\gamma$. Comparing this picture with Fig. \ref{fig:z35}, we notice a higher amplitude of the emerging helical loop, see Fig. \ref{fig:biforcazione_L12}. Moreover, in Fig. \ref{fig:ener_u0_L12} we observe that the energy lowers continuously, with a little delay on the onset of the instability due to the imposed boundary condition, and in Fig. \ref{fig:biforcazione_L12} the bifurcation diagram exhibits a continuous transition from the unbuckled to the buckled configuration, confirming the typical behaviour of a supercritical pitchfork bifurcation.

\begin{figure}[h!]
	\centering
	\includegraphics[width=0.5\textwidth]{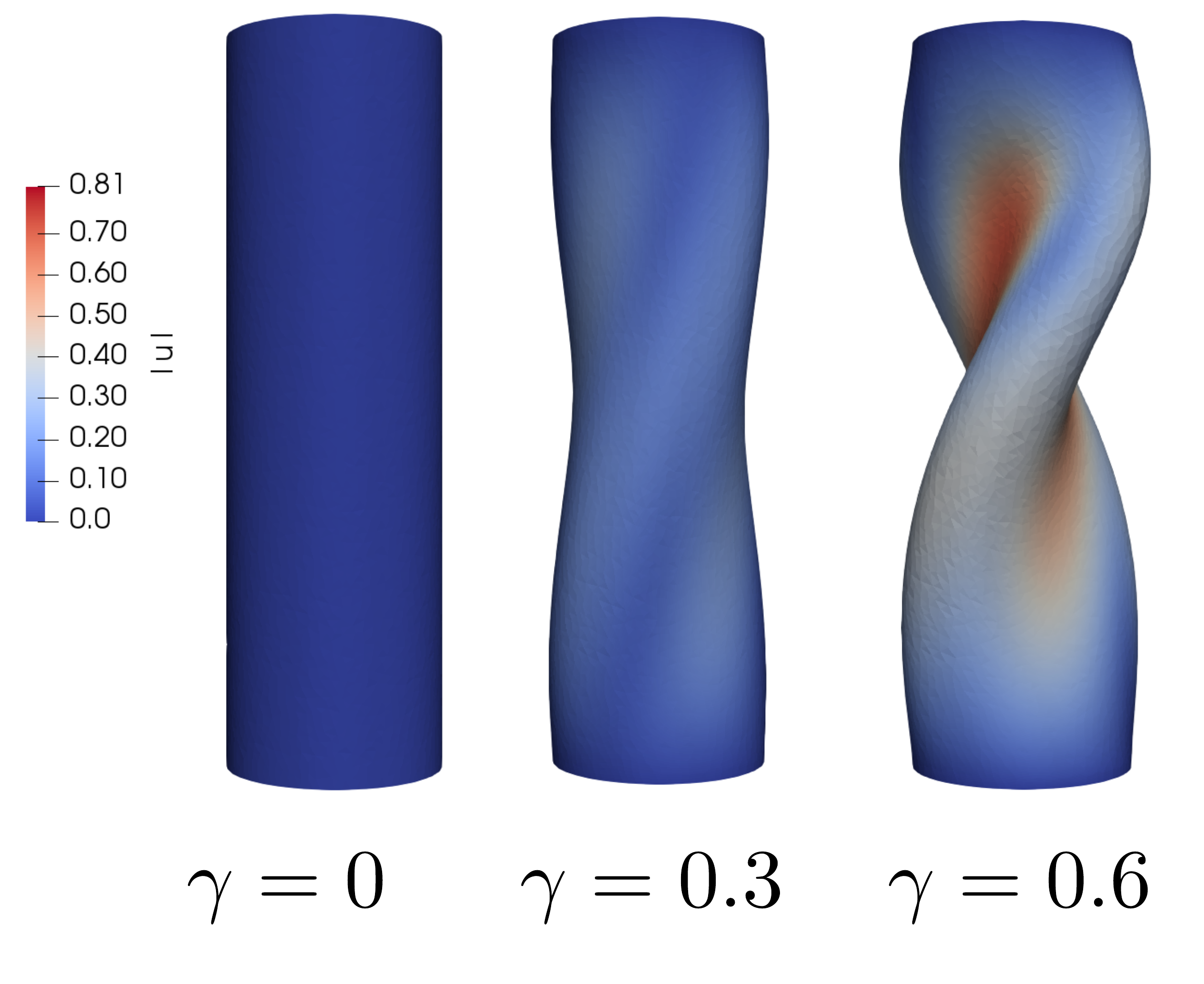}
	\caption{Actual configuration of the buckled tube when $\alpha_R = 1.35$ and $\alpha_L = 7$. In such conditions $m_{\rm cr} = 2$, $\tilde{k}_{\rm cr} = 2\pi /\alpha_L$ and $\gamma_{\rm cr}\simeq0.177001$.}
	\label{fig:sim074}
\end{figure}

Second, keeping fixed $\alpha_L = 7$, we consider a different initial aspect ratio $\alpha_R = 1.35$ to visualize the change in the mechanical response of the hollow cylinder. Imposing the boundary condition Eq. \eqref{eq:bcbasi_u0} on the two bases,  we plot in Fig. \ref{fig:sim074} the resulting buckled configuration. As found in the linear stability analysis,  the critical circumferential number is $m_{\rm cr} = 2$, while the critical axial wavenumber is always $2 \pi/\alpha_L$. The different pattern can be immediately visualized by comparing Fig. \ref{fig:sim285} with Fig. \ref{fig:sim074}, where the helical pattern exhibits a decreased amplitude.

\begin{figure}[h!]
	\centering
	\includegraphics[width=\textwidth]{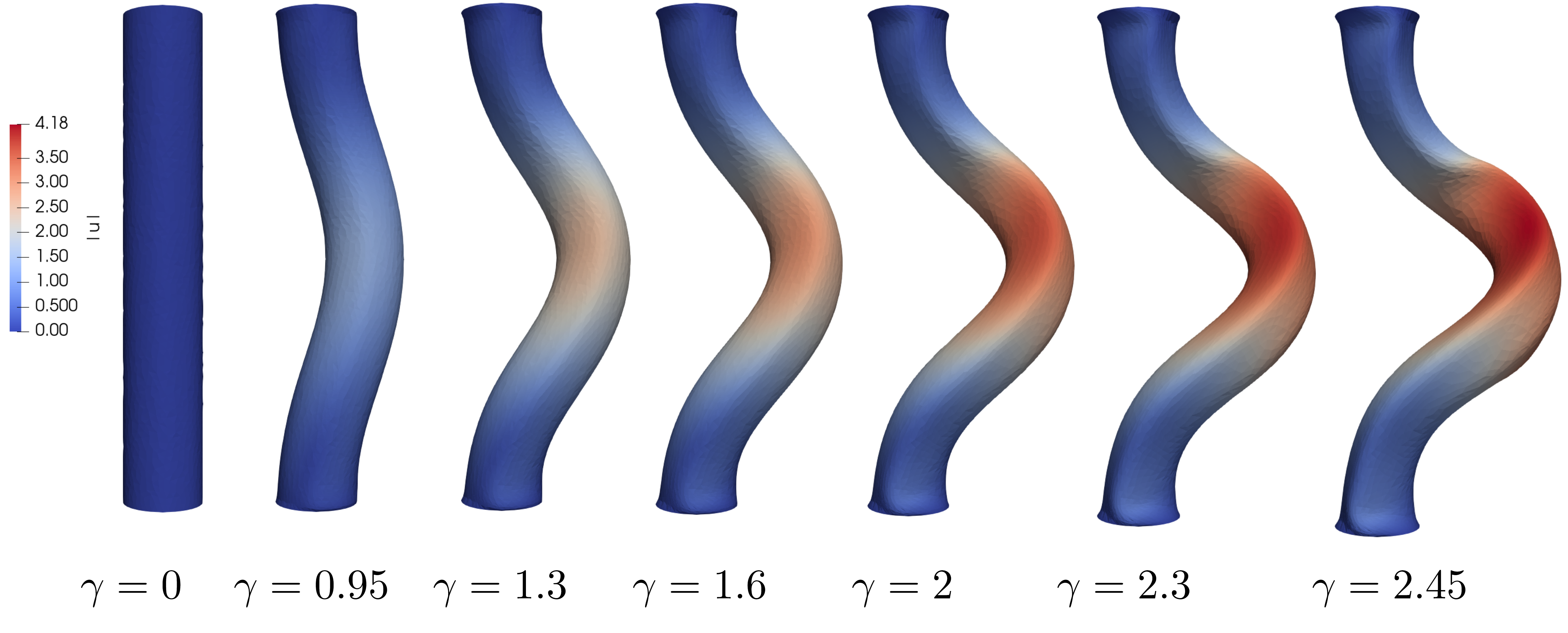}
	\caption{Actual configuration of the buckled tube for when $\alpha_R \to \infty$ and $\alpha_L = 12$. We choose $m_{\rm cr} = 1$ and $\tilde{k}_{\rm cr} = 2 \pi / \alpha_L$.}
	\label{fig:Ri0}
\end{figure}

Finally, if we consider a solid cylinder, that is characterized by $m_{\rm cr} = 1$, we validate our numerical results against  the classical problem of a twisted Euler rod, forming an helix of pitch $1/\tilde{k}$. For a Neo-Hookean material, this solution has been first presented by Green and Spencer in \cite{green1958stability}. Then, Gent and Hua \cite{gent2004torsional} investigated the evolution of this instability: it can evolve with the sudden onset of a sharply bent ring, or knot. Up to our knowledge, this paper is the first one which aims at reproducing the $3$D numerical simulation of a finite torsion rate on a soft solid cylinder. Fixing the geometry, $\alpha_L = 12$, $\alpha_R = +\infty$, since $R_{\rm i} = 0$, imposing the boundary condition Eq. \eqref{eq:bcbasi_u0} on the two bases,  we plot in Fig. \ref{fig:Ri0} the buckled configuration of the full cylinder. We show the actual configuration for several values of the control parameter $\gamma$. As $\gamma$ increases, as the cylinder displays an helical pattern with a higher amplitude compared to the one of hollow cylinders, we can compare Fig: \ref{fig:sim285} and Fig. \ref{fig:L12} with Fig. \ref{fig:Ri0}. The simulation stops around $\gamma \simeq 2.45$ probably due to the excessive distortion of the elements. Unfortunately, our simulation does not display the expected knot, since  we should consider a longer cylinder which, however, requires a fine mesh, hence a bigger computational effort. Future efforts will be devoted in improving the performance of the numerical simulations, taking also into account for the self-contact.

\begin{figure}[h!]
	\begin{subfigure}{.42\linewidth}
		\centering
		\includegraphics[width=0.8\textwidth]{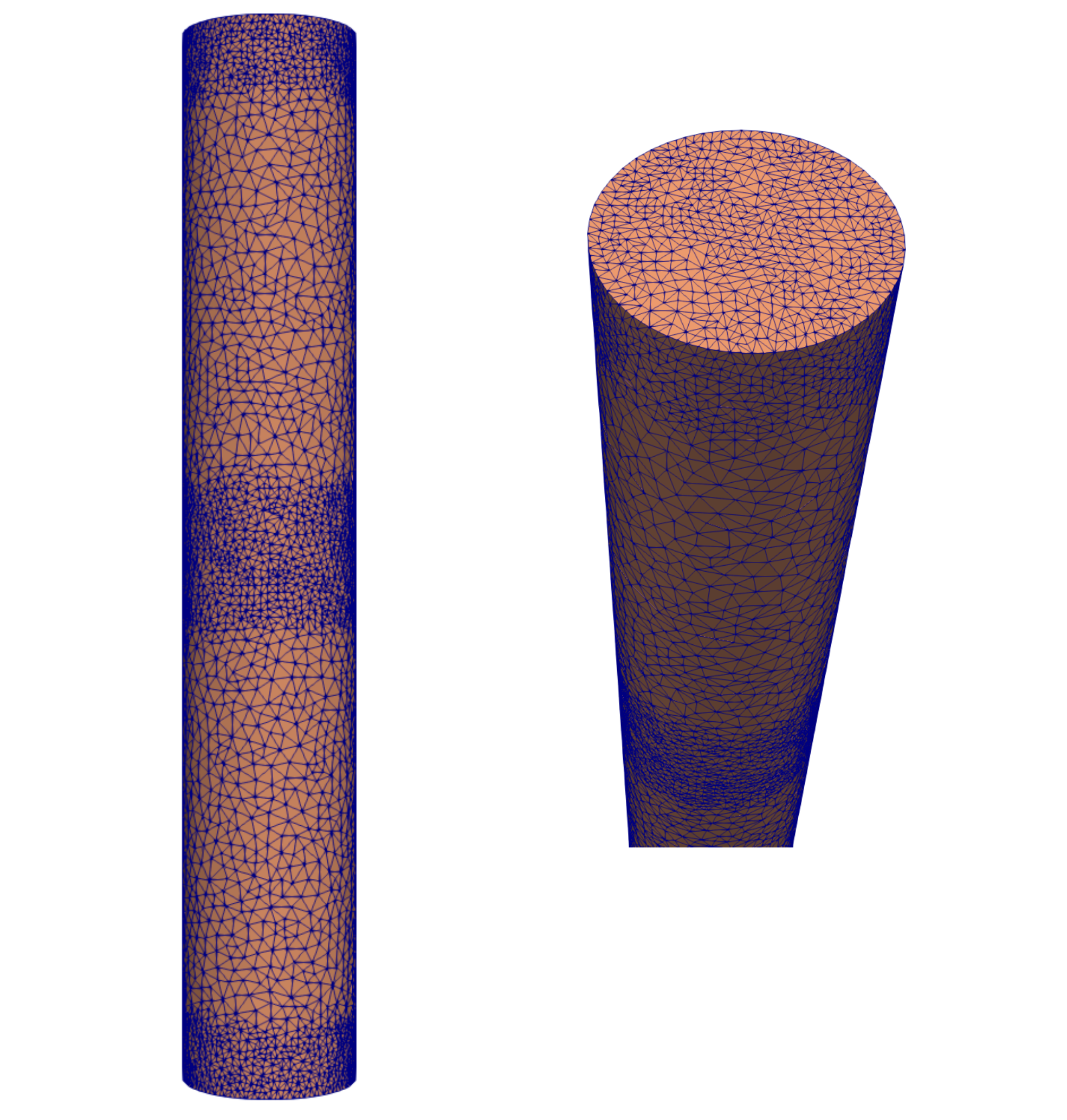}
		\caption{$\alpha_R \to \infty$}
		\label{fig:mesh_Ri0}
	\end{subfigure}
	\begin{subfigure}{.42\linewidth}
		\centering
		\includegraphics[width=0.7\textwidth]{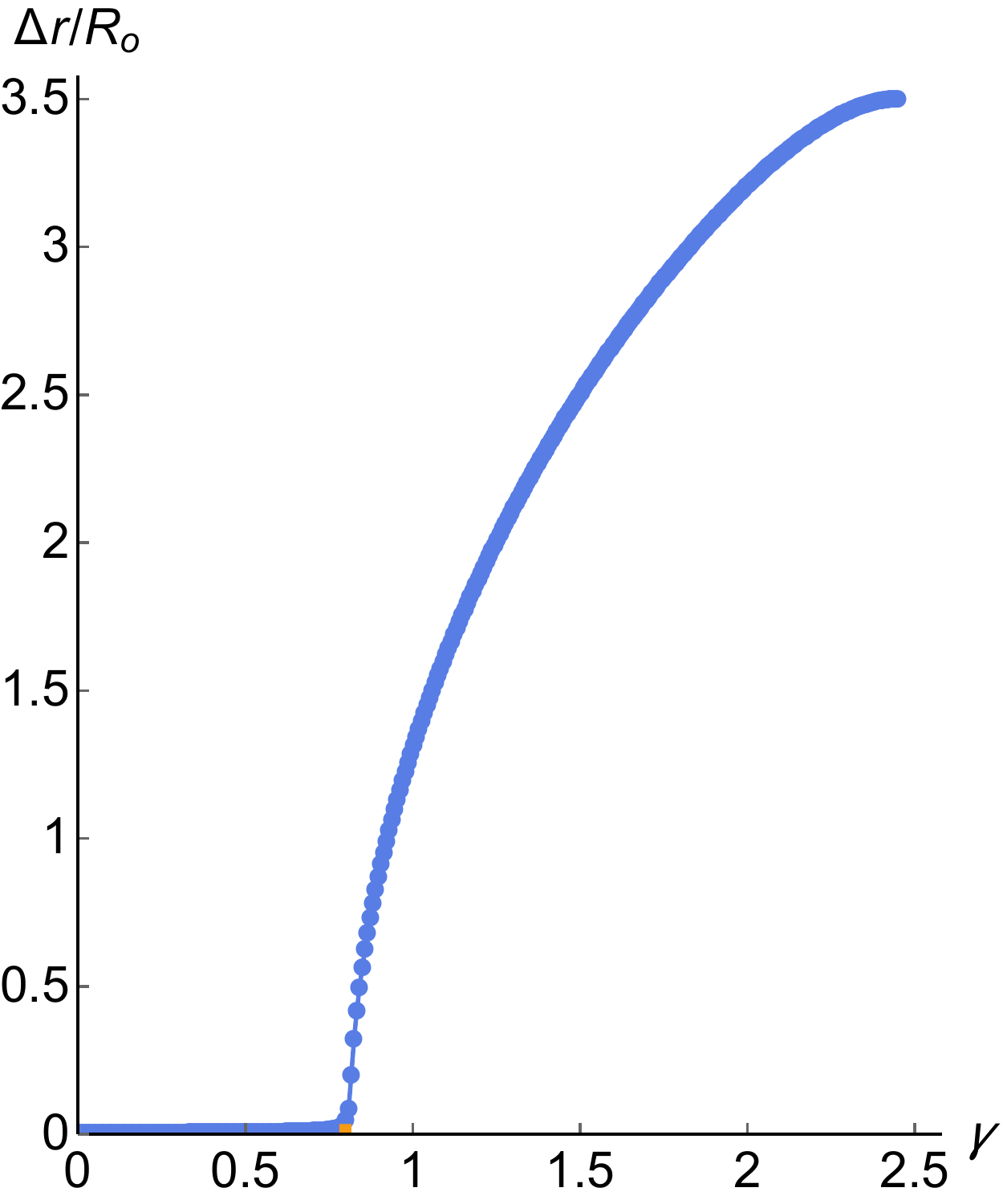}
		\caption{$\alpha_R \to \infty$ and $z = \alpha_L/2 = 6$}
		\label{fig:Ri0_nonlineare}
	\end{subfigure}%
	\caption{(a) Mesh generated through MSHR for $\alpha_R \to \infty$ with local refinements. (b) Bifurcation diagram showing the dimensionless parameter $\Delta r/ R_{\rm o}$ versus the control parameter $\gamma$ when $\alpha_R \to \infty$, $\alpha_L = 12$ and $z = \alpha_L/2 = 6$. The orange square denotes the marginal stability threshold $\gamma_{\rm cr} \simeq 0.8$ \cite{green1958stability}, with $m_{\rm cr} = 1$ and $\tilde{k}_{\rm cr} = 2\pi/\alpha_L$.}
	\label{fig:res_Ri0}
\end{figure}

In order  to reduce the distortion of the tetrahedra, we refine the mesh near the two bases and around $z = \alpha_L/2$, where we noticed the greatest concentration of elongated elements. The used mesh is presented in Fig. \ref{fig:mesh_Ri0}.
The linear stability analysis performed in Section \ref{sec:lin_stab_clooping} is no longer valid for a solid cylinder since the structure of the Stroh formulation changes in the absence of the inner lumen \cite{Ciarletta_2014}. In order to assess our numerical results, we compare the marginal stability threshold obtained from our numerical simulations with the theoretical one computed by Green and Spencer \cite{green1958stability}. In particular, we also found that the critical torsion is around $\gamma=0.8$  for $\tilde{k} = 2 \pi /\alpha_L$. In Fig. \ref{fig:Ri0_nonlineare}, we plot the distance of the centroid of the mid-section, precisely the quantity $\Delta r/R_{\rm 0}$ defined in Eq. \ref{eq:delta_r} versus the control parameter $\gamma$.

\section{Conclusions}
\label{sec:conclusion_clooping}
In this work, we have developed a simple morphomechanical model  to describe the dextral torsion during the c-looping of the HT, which represents the first asymmetry during the embryogenesis of the human heart. In Section \ref{sec:elastic_model_clooping}, we have proposed a nonlinear  elastic model of the HT undergoing torsional remodelling. The HT is described as a hollow cylinder whose aspect ratio and length are obtained by experimental observations \cite{taber2010role}.  The elastic BVP is governed by the dimensionless parameters $\gamma$, i.e. the finite torsion rate induced by the remodelling cell flow, $\alpha_R$, the ratio between outer and inner radius,  and $\alpha_L$, the ratio of the  length over the outer radius of the HT. We have computed a radially symmetric solution and we have studied its linear stability in Section \ref{sec:lin_stab_clooping} using the theory of incremental deformations \cite{ogden1997non}. We have rewritten the linear stability analysis into an optimal Hamiltonian system using the Stroh formulation, following a procedure similar to that proposed in \cite{balbi2015helical}.

The marginal stability thresholds are discussed in Section \ref{subsec:resultt_linstab}. Both the critical circumferential and axial modes strongly depend on the geometrical parameters $\alpha_R$ and $\alpha_L$. In particular, we recover some known results in the limit of thin tubes, as shown in Fig. \ref{fig:mvsalphaR}, showing a cut-off thickness at which the circumferential critical mode of the HT completely changes.  We also  highlight that increasing $\alpha_L$ lowers the instability threshold  \cite{green1958stability,gent2004torsional}, as shown in Fig. \ref{fig:vsalphaL_mcr1}. 

Finally,  we have implemented in Section \ref{sec:numerical_simulation_clooping} a finite element code to approximate the fully non-linear BVP. We use a mixed variational formulation whose linearization is based on the Newton method. The outcomes of our numerical simulations are reported in Figs. \ref{fig:sim285} - \ref{fig:res_Ri0}. We have considered both a physiological range of the geometrical parameters with experimental data, see Figs. \ref{fig:sim285} - \ref{fig:ener}, different geometrical data to validate our code, see Figs. \ref{fig:alphaL12} - \ref{fig:sim074} and the solid cylinder limit, see Figs. \ref{fig:Ri0} - \ref{fig:res_Ri0}. These results show how the geometry of the cylinder, both the thickness and the slenderness ratio, strongly affect the looping onset  and its  nonlinear development. In all the cases, the bifurcation is supercritical, displaying a continuous transition from the axis-symmetric to the looped configurations, see Figs. \ref{fig:nonlineare}, \ref{fig:ener}, \ref{fig:L12} and \ref{fig:Ri0_nonlineare}.  We finally performed a grid convergence analysis showing that our numerical results do not change any longer if a further mesh refinement is operated, as shown in  Fig. \ref{fig:conv_analisis}.

In conclusion, our simple morphomechanical model suggests that a torsional internal remodelling alone can drive the spontaneous onset and the fully nonlinear development of the c-looping of the HT within its  physiological range of geometrical parameters.  This works aims to prove that mechanical features are as important as  biological and chemical processes during this stage of heart embryogenesis.   Further developments  will be directed to investigate if the symmetry break results from the cell flow remodelling or  may be directed from external constraints.  We also aim to perform numerical simulations using a more realistic geometry extracted from bioimaging data.

\section*{Acknowledgements}
We wish to thank Davide Ambrosi, Luca Ded\'e, Simone Pezzuto and Davide Riccobelli for helpful suggestions and fruitful discussions.\\
GB nd PC acknowledge the support from MIUR, PRIN 2017 Research Project ”Mathematics of active materials: from mechanobiology to smart devices”. The work of GB and PC is partially supported by GNFM-INdAM. AQ has received funding from the European Research Council (ERC) under the European Union’s Horizon 2020 research and innovation programme (grant agreement No 740132, iHEART - An Integrated Heart Model for the simulation of the cardiac function, P.I. Prof. A. Quarteroni).
\newpage

\bibliographystyle{abbrvnat}
\bibliography{refs}
\end{document}